\documentclass[10pt, 3p]{elsarticle}

\usepackage[british]{babel}
\usepackage[utf8]{inputenc}
\usepackage{graphics}
\usepackage{soul}
\usepackage{subcaption}
\usepackage{graphicx}
\usepackage{amsmath,amssymb,esint}
\usepackage{lineno}
\usepackage{multirow}
\usepackage{natbib}
\usepackage{eucal}
\usepackage{mathtools}
\usepackage{placeins}
\usepackage{enumerate}
\usepackage[usenames,dvipsnames]{xcolor}
\usepackage[colorlinks]{hyperref}
\usepackage{setspace}
\usepackage{MnSymbol}
\usepackage{siunitx}

\biboptions{sort&compress}
\journal{Journal of Computational Physics}

\begin{document}
\newcommand{\reviewer}[1]{{{#1}}}
\newcommand{\reviewerRevtwo}[1]{{\color{black}{#1}}}
\newcommand{\Editor}[1]{{\color{black}{#1}}}

\begin{frontmatter}

\title{Sharp front tracking with geometric interface reconstruction}

\author[affil1]{Christian Gorges}
\author[affil2]{Fabien Evrard}
\author[affil4]{Robert Chiodi}
\author[affil1]{Berend van Wachem}
\author[affil5]{Fabian Denner\corref{cor1}}
\ead{fabian.denner@polymtl.ca}

\address[affil1]{Chair of Mechanical Process Engineering,
Otto-von-Guericke-Universit\"{a}t Magdeburg,\\ Universit\"atsplatz 2, 39106 Magdeburg, Germany}
\address[affil2]{Department of Aerospace Engineering, University of Illinois Urbana-Champaign,\\ Urbana, IL, 61820, United States of America}
\address[affil4]{Computer, Computational, and Statistical Sciences Division, Los Alamos National Laboratory,\\ Los Alamos, NM, 87545, United States of America}
\address[affil5]{Department of Mechanical Engineering, Polytechnique Montr\'eal,\\ Montr\'eal, QC, H3T 1J4, Canada}

\cortext[cor1]{Corresponding author: }

\begin{abstract}
\reviewer{This paper presents a novel sharp front-tracking method designed to address limitations in classical front-tracking approaches, specifically their reliance on smooth interpolation kernels and extended stencils for coupling the front and fluid mesh. In contrast, the proposed method employs exclusively sharp, localized interpolation and spreading kernels, restricting the coupling to the interfacial fluid cells—those containing the interface/front. This localized coupling is achieved by integrating a divergence-preserving velocity interpolation method with a piecewise parabolic interface calculation (PPIC) and a polyhedron intersection algorithm to compute the indicator function and local interface curvature. Surface tension is computed using the \reviewerRevtwo{Continuum} Surface Force (CSF) method, maintaining consistency with the sharp representation. Additionally, we propose an efficient local roughness smoothing implementation to account for surface mesh undulations, which is easily applicable to any triangulated surface mesh. Building on our previous work, the primary innovation of this study lies in the localization of the coupling for both the indicator function and surface tension calculations. By reducing the interface thickness on the fluid mesh to a single cell, as opposed to the 4–5 cell spans typical in classical methods, the proposed sharp front-tracking method achieves a highly localized and accurate representation of the interface. This sharper representation mitigates parasitic currents and improves force balancing, making it particularly suitable for scenarios where the interface plays a critical role, such as microfluidics, fluid-fluid interactions, and fluid-structure interactions. The proposed method is comprehensively validated and tested on canonical interfacial flow problems, including stationary and translating Laplace equilibria, oscillating droplets, and rising bubbles. The presented results demonstrate that the sharp front-tracking method significantly outperforms the classical approach in terms of accuracy, stability, and computational efficiency. Notably, parasitic currents are reduced by approximately two orders of magnitude and stable results are obtained for parameter ranges where classical front tracking fails to converge.}
\end{abstract}
\end{frontmatter}

\section{Introduction}

\begin{figure}
    \centering
    \subcaptionbox{Classic front tracking\label{fig:Indicator_Classic}}{\includegraphics[width=6cm]{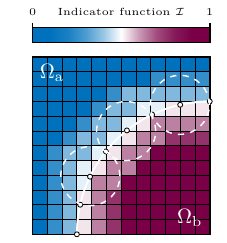}}
	\subcaptionbox{Sharp front tracking\label{fig:Indicator_Sharp}}{\includegraphics[width=6cm]{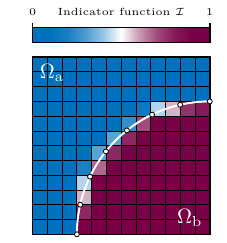}}\\
    \caption{Qualitative illustration of the indicator function representation on the Eulerian fluid mesh. Due to the large spreading stencils, the classic front tracking (a) leads to a smeared out interface representation on the Eulerian fluid mesh, whereas the geometric interface reconstruction of the sharp front tracking (b) leads to a sharp interface representation on the Eulerian fluid mesh.}
    \label{fig:IndicatorComparison}
\end{figure}

Front tracking \cite{Unverdi1992, Tryggvason2001} is a widely-used method to model immiscible interfacial flows. In the three-dimensional front-tracking (FT) method, the interface is represented by a triangulated surface mesh, of which the vertices are advected in a Lagrangian manner by the underlying flow. The surface mesh enables a straightforward computation of the interface properties \cite{Tryggvason2011a}, e.g.~the surface tension, and provides a distinct computational domain for surface-adsorbed substances \cite{Muradoglu2008,Muradoglu2014}. A critical step in front tracking is sharing information between the fluid mesh and the surface mesh, such as the fluid velocity, the position of the surface mesh to define the fluid properties and the force due to surface tension, which requires careful consideration. In moving flows or when the interface is evolving, the fluid mesh and the surface mesh do not match in terms of the position of their vertices and, in turn, information shared between the meshes has to be interpolated. Typically, smooth kernels with a compact support of multiple mesh cells, such as the cosine kernel of \citet{Peskin1977}, are used for the interpolation between the fluid mesh and the surface mesh \cite{Tryggvason2011a}, which leads to a smeared-out representation of the interface that depends on a length scale associated with the interpolation kernels (see Figure \ref{fig:Indicator_Classic}). Consequently, it is widely recognized \cite{Tryggvason2001} that the support of the interpolation kernel should be as narrow as possible without introducing an undue mesh dependence.

In order to distinguish the interacting fluid phases, an indicator function is reconstructed on the fluid mesh based on the position of the surface mesh. Reconstructing this indicator function is typically done in three steps: defining the indicator gradient on the surface mesh, i.e.~the interfacial normal vector, interpolating the indicator gradient from the front mesh to the fluid mesh using a smooth interpolation kernel, and reconstructing the indicator function by solving a Poisson problem. This procedure yields an indicator function and, consequently, fluid properties that vary over several mesh cells at the interface. Alternatives to reconstructing the indicator function have been based on closest-point transforms \cite{Ceniceros2005}, and hybrid FT/Volume-of-Fluid \cite{Aulisa2003a} or FT/level-set methods \cite{Popinet1999, Shin2018, Tolle2020}. Similarly, the force on the fluid due to surface tension is computed directly on the surface mesh, exploiting the explicit representation of the fluid interface, and then interpolated to the fluid mesh, again using a smooth kernel. As a result, the common way of reconstructing the indicator function as well as the interpolation of the force due to surface tension both yield a smeared-out representation of the interface. 

When interpolating the fluid velocity from the fluid mesh to the surface mesh, where it is used to transport the surface mesh with the flow, a similar problem arises. The use of smooth interpolation kernels means that the motion of the surface mesh depends on a large neighbourhood and the divergence of the velocity field may not be preserved upon interpolation, which leads to mass conservation errors \cite{Takeuchi2020}. An alternative to smooth interpolation kernels is the frequently used trilinear interpolation of the velocity from the neighbouring cells of the fluid mesh \cite{Tryggvason2011a}. While this reduces the stencil size of the interpolation, it does not include any provisions to actively preserve the divergence of the velocity field upon interpolation, resulting in mass conservation errors similar to smooth interpolation kernels \cite{Gorges2022}. In an effort to retain a small interpolation stencil and minimize conservation errors, we have recently proposed a divergence-preserving velocity interpolation method that reduces conservation errors by about one order of magnitude \cite{Gorges2022}.

The spread-out interface representation resulting from using relatively large interpolation stencils presents important practical limitations to FT methods in relation to predicting the dynamic behaviour of relatively poorly resolved fluid structures (e.g.~droplets, bubble, ligaments), as well as the dynamics of interfaces in close proximity to each other or a wall.
In addition, for interfacial flows in which surface tension dominates, FT methods routinely fail at achieving the Laplace equilibrium of a circular or spherical interface by reducing parasitic currents to solver tolerance or machine precision, as a result of applying a large interpolation stencil for both the velocity and the force due to surface tension.
While some alternatives to smooth interpolation kernels for velocity interpolation or interface representation have been proposed \cite{Tryggvason2011a, Gorges2022, Dijkhuizen2010, Tavares2021} for the particular challenges of front tracking, a sharp FT method that only uses local information and does not require smooth interpolation kernels for all unit operations is not available to date.

Because of numerical errors in the velocity interpolation and variations in the velocity field from one fluid cell to the next, or in regions where the front vertices tend to accumulate due to underlying flow field, undulations may occur and introduce a local roughness on the front mesh.
For instance, this phenomenon typically occurs at the bottom of rising bubbles or in regions with strong parasitic currents arising from imbalances in the pressure gradient and surface tension forces \cite{deSousa2004}.
\reviewer{Such undulations are typically much smaller than the fluid mesh spacing and are not present in real flows, as they are physically removed by a combination of surface tension and viscous effects. A coupling between flow and front that acts at the cell level cannot correctly suppress these front undulations. This non-physical behaviour introduces errors in the computation of the geometric front properties and is accompanied by a degradation of an accurate front advection and surface tension computation. The occurrence of such spurious undulations and local surface roughness is one reason why a fluid mesh spacing to front mesh spacing ratio of $\Delta x / l_e \approx 1$ is required to achieve numerical stability in FT simulations \citep{Tryggvason2011a}. To the best of our knowledge, every $\Delta x / l_e$ ratio despite $\Delta x / l_e \approx 1$ should be used with caution. $\Delta x / l_e > 1$ is not helpful because information is lost during spreading to a coarser fluid mesh and only leads to unnecessary computational overhead. 
Sub-cell undulations and parasitic currents become more problematic as $\Delta x / l_e$ decreases and roughness smoothing becomes essential, since  $\Delta x / l_e < 1$ typically leads to an inaccurate interface representation, alongside an inaccurate velocity interpolation and surface tension spreading.}
Many methods to suppress these non-physical undulations and remove surface roughness on surface meshes have been developed in the context of computer graphics and computer-aided design, for instance the use of a Gaussian filter, Laplacian smoothing and its variations, curvature and diffusion flow methods \cite{Jiao2006, Desbrun1999, Dunyach2013}. However, in front tracking, it is important for the technique applied not to alter the volume enclosed by the front mesh, which makes Laplacian smoothing, curvature and diffusion flow methods impracticable. For FT simulations, the TSUR-3D method \cite{deSousa2004} is a frequently used method to smooth small front undulations and retain the local volume enclosed by the front. However, the user must decide \textit{a priori} when the method is to be applied and the method is applied to the entire front mesh and not just to a small, localized area in which the front mesh has undulations or localized roughness. If the TSUR-3D algorithm is applied too frequently, the general shape of the interface may be distorted. 

\reviewer{In this article, we propose a \textit{sharp} front-tracking method, which eliminates the need for smooth kernels commonly used in classical front-tracking methods (see Figure \ref{fig:Indicator_Sharp}). Unlike traditional front tracking approaches that rely on extended interpolation stencils spanning multiple fluid mesh cells, the proposed method restricts the coupling between the interface and fluid mesh to the interfacial fluid cells — those cells directly containing the interface. This sharp coupling is achieved using a compact divergence-preserving velocity interpolation method from our previous work \cite{Gorges2022}. In comparison to our earlier work, the primary innovation of the present study is the localization of the coupling for both the indicator function and surface tension calculations. To achieve this, we integrate a piecewise parabolic interface calculation (PPIC) with a polyhedron intersection algorithm \cite{Chiodi2022, Evrard2023} to compute the curvature and indicator function exclusively within the interfacial fluid cells. To the best of our knowledge, such a sharp and localized coupling strategy has not been applied in front tracking previously. Surface tension is computed using the \reviewerRevtwo{Continuum} Surface Force (CSF) method \cite{Brackbill1992}, maintaining consistency with the sharp representation. The choice of the CSF method for the surface tension force is motivated primarily by the fact that it can provide an accurate force balance, as demonstrated exhaustively in the literature with VOF methods \citep{Popinet2018,Janodet2025}. Moreover, the CSF method can be readily deployed with PPIC, in conjunction with the polyhedron intersection algorithm, for the accurate communication and calculation of the interfacial properties and volume fraction in the interfacial fluid mesh cells. Nonetheless, the methods used in this work are not the only possible choices and other methods (e.g.~the height function method for the computation of the interface curvature) could be considered as well.

The term "sharp" refers to both the interfacial representation on the fluid mesh and the stencil used for coupling the front and fluid mesh. In classical front tracking, the interface thickness on the fluid mesh spans approximately 4-5 mesh cells, due to the Peskin spreading and interpolation. In contrast, the proposed sharp front-tracking method reduces the interface thickness to a single cell, achieving a far more localized representation of the interface on the fluid mesh.
We combine this sharp front-tracking method with a local roughness smoothing that is tailored to front-tracking methods, in order to reduce the local surface roughness and undulations. Using a state-of-the-art finite-volume algorithm, the sharp front-tracking method is shown to yield a clear improvement of the interface representation on the fluid mesh and a reduction of parasitic currents by up to two orders of magnitude compared to the classic front-tracking method \cite{Tryggvason2001}.
}

This article is structured as follows: In section \ref{sec:governing}, the numerical framework, in which the sharp front-tracking method is embedded, is described. In section \ref{sec:numerics}, the sharp front-tracking method is then derived in detail, followed by section \ref{sec:results}, in which the proposed method is verified, validated and compared against the classic front-tracking method for canonical interfacial flow test cases. Section \ref{sec:conclusions} provides a final summary and conclusion.

\section{Governing equations and numerical framework}
\label{sec:governing}

The considered incompressible interfacial flows are governed by the continuity and momentum equations, given in the one-fluid formulation as
\begin{align}
  \nabla \cdot \mathbf{u} &= 0 \label{eq:continuity} \\
  \rho \left[ \frac{\partial \mathbf{u}}{\partial t} +  \nabla \cdot (\mathbf{u} \otimes \mathbf{u}) \right] &= - \nabla p + \nabla \cdot \left[\mu \left(\nabla \mathbf{u} + \nabla \mathbf{u}^\mathrm{T} \right)\right] + \mathbf{S}, \label{eq:momentum}
\end{align}
respectively, where $t$ denotes time, $\mathbf{u}\equiv \begin{bmatrix} u & v & w\end{bmatrix}^\intercal$ is the velocity vector, $p$ the pressure, $\rho$ is the  density, $\mu$ is the viscosity, and $\mathbf{S}$ is the sum of volumetric momentum source terms applied to the flow. This study considers the source terms representing the force on the fluid arising from surface tension, $\mathbf{S}_\sigma$, and gravity, $\mathbf{S}_g$, such that  $\mathbf{S}=\mathbf{S}_\sigma+\mathbf{S}_g$. The discretization of $\mathbf{S}_g$ is detailed in \cite{Denner2014a}. The immiscible fluids are distinguished by an indicator function $\mathcal{I}$ that is reconstructed based on the position of the interface and is defined as
\begin{equation}
  \mathcal{I}(\mathbf{x}) = \begin{cases}
    0, &\mbox{if} \ \mathbf{x} \in \Omega_\mathrm{a}\\
    1, &\mbox{if} \ \mathbf{x} \in \Omega_\mathrm{b}
  \end{cases}
\end{equation} 
where $\Omega_\text{a}$ and $\Omega_\text{b}$ are the subdomains occupied by fluids ``a'' and ``b'', respectively. The density and viscosity are then given as
\begin{align}
  \rho(\mathrm{x}) &= \rho_\mathrm{a} + \mathcal{I}(\mathbf{x}) (\rho_\mathrm{b}-\rho_\mathrm{a})\\
  \mu(\mathrm{x}) &= \mu_\mathrm{a} + \mathcal{I}(\mathbf{x}) (\mu_\mathrm{b}-\mu_\mathrm{a}).
\end{align}

\reviewer{As a basis for the proposed sharp FT method, the fully-coupled pressure-based algorithm of \citet{Denner2020} is used. In this algorithm, the governing equations are discretized using a second-order finite-volume discretization for the advection, pressure gradient and stress tensor, and a second-order backward Euler scheme for the transient term. The equations are solved implicitly in a single linear system of equations for pressure and velocity.} All fluid variables are stored in a collocated variable arrangement at the cell centers of the fluid mesh. The required pressure-velocity coupling is achieved by applying a momentum-weighted interpolation (MWI) \cite{Rhie1983} to define the fluxes through the mesh faces. Assuming an equidistant Cartesian mesh, including source terms and accounting for large density ratios, the flux through mesh face $f$ reads as \cite{Bartholomew2018}
\begin{multline}
  F_f = \left\{ \overline{\mathbf{u}}_{f} \cdot \mathbf{n}_{f} - \hat{d}_f \left[ \frac{p_Q-p_P}{\Delta x} -  \frac{\rho^\ast_f}{2} \left(\frac{\nabla p_P}{\rho_P} + \frac{\nabla p_Q}{\rho_Q} \right) \cdot \mathbf{n}_{f} \right] +
   \hat{d}_f \left[\mathbf{S}_f \cdot \mathbf{n}_f - \frac{\rho^\ast_f}{2} \left(\frac{\mathbf{S}_P}{\rho_P} + \frac{\mathbf{S}_Q}{\rho_Q} \right) \cdot \mathbf{n}_{f} \right] \right. \\
  + \left. \hat{d}_f \frac{\rho^\ast_f}{\Delta t} \left( \vartheta_f^{(t-\Delta t)} - \overline{\mathbf{u}}_{f}^{(t-\Delta t)} \cdot \mathbf{n}_{f} \right)  \right\} A_f,
  \label{eq:mwi}
\end{multline}
where $P$ and $Q$ are the two mesh cells adjacent to face $f$, $\rho_f^\ast$ is the density at face $f$ obtained by harmonic averaging of the densities at the adjacent mesh cells, and $A_f$ is the area of face $f$. The weighting factor $\hat{d}_f$ is defined by the discretization of the advection and shear stress terms of the momentum equations, as described in detail by \citet{Bartholomew2018}.
\reviewer{For more information on the details of the numerical framework, the interested reader is referred to the work of Denner et al.~\citep{Denner2014a,Denner2020}.}

\section{Sharp front tracking}
\label{sec:numerics}

In this section a detailed explanation of the modelling and algorithms of the sharp FT method is given and is contrasted with the classic FT method.

\subsection{Velocity interpolation}

The vertices of the surface mesh are advected in a Lagrangian manner using a fourth-order Runge-Kutta scheme \cite{Gorges2022} to discretize the ordinary differential equation
\begin{equation}
   \reviewerRevtwo{ \frac{\mathrm{d} \mathbf{x}_i(t)}{\mathrm{d}t} = \bar{\mathbf{u}}(t, \mathbf{x}_i).}
\end{equation}
where $\mathbf{x}_i$ and $\bar{\mathbf{u}}$ are the location and (interpolated) velocity of vertex $i$, respectively. Since the velocity is only known at the cell centers and faces of the fluid mesh, the velocity $\bar{\mathbf{u}}(t, \mathbf{x}_i)$ of the vertex of the surface mesh has to be interpolated from the fluid mesh. Classically, Peskin interpolation \cite{Peskin1977} is used in front tracking \cite{deJesus2015,Hua2008, Pivello2012, Pivello2014, Gorges2023}, whereby the cell- or face-centered velocities are interpolated within a symmetrical stencil of twice the fluid mesh spacing in each direction using the cosine function
\begin{equation}
d(r) = 
\begin{cases}
\dfrac{1}{4} \left[ 1+ \mathrm{cos} \left( \dfrac{\pi r}{2} \right) \right], &  \mathrm{if} \ |r|<2 \\
0, &  \mathrm{if} \ |r|\geq 2, 
\end{cases} \label{eq:peskin}
\end{equation}
where $r_x = (x_i - x_c)/\Delta x$ is the distance between the $x$-position of the Lagrangian vertex $i$ and the fluid mesh cell center $c$, normalized by the fluid mesh spacing $\Delta x$. This leads to a discrete interpolation operator that is the product of three one-dimensional (1D) interpolation kernels
\begin{equation}
    \bar{\mathbf{u}}(x_i,y_i,z_i) = \sum_{c=1}^{64} \mathbf{u}(\mathbf{x}_c) w(\mathbf{x}_c),
\end{equation}
with $w(\mathbf{x}_c) = d(r_x) d(r_y) d(r_z)$, and that uses 64 fluid mesh velocity points (in three dimensions) around the target Lagrangian vertex. A qualitative illustration of the Peskin interpolation stencil in two dimensions is also shown in Figure \ref{fig:Indicator_Classic}, where the white dashed circles represent the respective stencil for their centered Lagrangian vertices.

In contrast to the commonly used Peskin interpolation, the proposed sharp FT method applies the divergence-preserving velocity interpolation method first introduced by us \cite{Gorges2022} for FT simulations and based on the work of \citet{Toth2002}. This method interpolates the velocity inside a cell as the combination of two terms: 1. the linear interpolation of velocity using the face velocities; 2. a correction based on transverse velocity gradients that preserves divergence inside the cell. 
For simplicity, considering an arbitrarily sized three-dimensional (3D) Cartesian fluid mesh cell ranging from $\smash{\begin{bmatrix}x^- & y^- & z^- \end{bmatrix}^\intercal}$ to $\smash{\begin{bmatrix}x^+ & y^+ & z^+ \end{bmatrix}^\intercal}$; the discrete normal face velocities at the cell faces are $U^{\pm}$, $V^{\pm}$ and $W^{\pm}$, where the superscripts denote the coordinates of the face centers. The interpolated velocity $\smash{\mathbf{\bar{u}} = \begin{bmatrix}\bar{u} & \bar{v} & \bar{w} \end{bmatrix}^\intercal}$  at the position $\smash{\begin{bmatrix}x & y & z \end{bmatrix}^\intercal}$ inside this arbitrarily sized Cartesian cell is given by the divergence-preserving interpolation polynomials as
\begin{equation}
    \begin{split}
       \bar{u}(x,y,z) & = 
    \frac{x - x^-}{\Delta x} \left(U^{+} + \left( y - \frac{y^- + y^+}{2} \right)U_y^{+} + \left( z - \frac{z^- + z^+}{2} \right)U_z^{+}\right)\\ 
    &+ \frac{x^+ - x}{\Delta x} \left(U^{-} + \left( y - \frac{y^- + y^+}{2} \right)U_y^{-} + \left( z - \frac{z^- + z^+}{2} \right)U_z^{-}\right)\\
    &+ \frac{(x^+ - x)(x - x^-)}{2\Delta x} \left(V_x^{+} - V_x^{-} + W_x^{+} - W_x^{-}\right),
    \end{split}
\end{equation}
$\bar{v}(x,y,z)$ and $\bar{w}(x,y,z)$ following the same ruling as $\bar{u}(x,y,z)$. These interpolation polynomials ensure that the discrete divergence of the underlying fluid velocity field is preserved \reviewer{(see \citet{Gorges2022} for a definition of the discrete divergence)}. The interpolation polynomials are continuous across cell faces in the normal direction, but discontinuous in the transverse direction, because the transverse gradients do not exactly match on both sides of a face connecting two cells \cite{Gorges2022, Toth2002}. If a Lagrangian vertex crosses the face between two cells, the derivatives of its trajectory may be subject to a discontinuity as the velocity changes abruptly which may lead to oscillations. Nevertheless, this discontinuity vanishes with increasing fluid mesh resolution with second order. A more detailed step-by-step derivation of the divergence preserving velocity interpolation and its components is given by \citet{Gorges2022}.

\subsection{Geometric interface reconstruction}

\begin{figure}
	\centering
	\subcaptionbox{\label{fig:PF_1}}{\includegraphics[width=5cm]{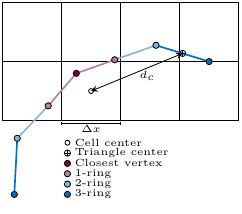}}
	\subcaptionbox{\label{fig:PF_2}}{\includegraphics[width=5cm]{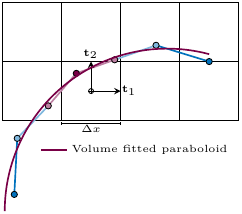}}
 \subcaptionbox{\label{fig:PF_3}}{\includegraphics[width=5cm]{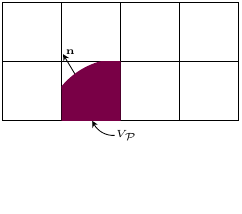}}\\
	\caption{A 2D illustration of the substeps of the parabolic interface reconstruction and polyhedron intersection algorithm. The coloured vertices represent the front mesh vertices and the coloured edges represent the front mesh triangles.}
	\label{fig:ParabolicFit}
\end{figure}

The discrete indicator function $\alpha$, also referred to as the volume fraction field, for a given fluid mesh cell $C$ is defined as
\begin{equation}
    \alpha_{C} = \frac{1}{V_C} \int_C \mathcal{I}(\mathbf{x}) \, \mathrm{d}x, 
\end{equation}
with $V_C$ being the volume of the fluid mesh cell. In the classic FT method, the discrete indicator function on the fluid mesh is computed as the solution of the Poisson equation \cite{Tryggvason2001} 
\begin{equation}
    \nabla^2 \alpha = \nabla \cdot \left( \sum\limits_{t} D(\mathbf{x} - \mathbf{x}_t) \mathbf{n}_t\right),
\end{equation}
whereby $\nabla \alpha = \sum_{t} D(\mathbf{x} - \mathbf{x}_t)\mathbf{n}_t $ is the indicator gradient spread from the surface mesh onto the fluid mesh using a distribution function $D$, for instance, the Peskin kernel, $\sum_{t}$ is the sum over all front triangles, and $\mathbf{n}_t$ is the triangle normal vector.

The proposed sharp front tracking method utilizes a piecewise-parabolic interface calculation (PPIC) to obtain the indicator function on the fluid mesh. For the interfacial cells and their neighbours, an implicit paraboloid surface is fitted to the surface mesh. Within a local orthonormal basis $\mathbf{\mathcal{T}} = \begin{bmatrix}\mathbf{t}_1 & \mathbf{t}_2 & \mathbf{n} \end{bmatrix}$, whose origin is the fluid cell center $\mathbf{x}_c$ (see Figure \ref{fig:ParabolicFit}), with $\mathbf{n}$ being the normal vector of the closest front-vertex, the implicit paraboloid surface $\mathcal{P}$ is defined in the parametric form
\begin{equation}
    z = \mathcal{P}(x,y) = a_1 + a_2 x + a_3 y + a_4 x^2 + a_5 xy + a_6 y^2.
\end{equation} 
The paraboloid fitting problem, i.e., finding the six coefficients $\smash{\mathbf{a}=\begin{bmatrix} a_1 & a_2 & a_3 & a_4 & a_5 & a_6 \end{bmatrix}^\intercal}$ that lead to an optimal local parabolic approximation of the interface, can be formulated as a least-squares minimization. Depending on the type of discrete surface elements that are considered, this can take two distinct forms: 1.~fitting the paraboloid to a set of neighbouring vertices by finding the minimum algebraic distance between the implicit surface and the Lagrangian vertex sample points (\textit{vertex fit}) \cite{Renardy2002, Gorges2022, Scardovelli2003, Han2024}, or 2.~minimising the volume between the front triangles in a local neighbourhood and the local paraboloid approximation (\textit{volume fit}) based on the work of \citet{Jibben2019}. Preliminary results have shown that for the purpose of the sharp FT method, the volume fit performs better than the vertex fit both in terms of stability and accuracy, and is therefore the method of choice in this work. This volumetric fitting was originally designed to minimize the volume between PLIC reconstruction polygons in a local neighbourhood of fluid mesh cells and the local paraboloid approximation. For our proposed FT method, the PLIC polygons used in \cite{Jibben2019} are replaced by the front triangles. In the local orthonormal basis $\mathbf{\mathcal{T}}$, the volume fit problem takes the form of the weighted least-squares minimization
\begin{equation}
    \min\limits_{\,\mathbf{a}\in\mathbb{R}^6}\ \sum_t w_t \left(\int_{\Gamma_t}\left(\mathcal{P}\left(x, y \right)-g_t \left(x, y \right)\right) \mathrm{d} A\right)^2,
    \label{eq:volumefit}
\end{equation}
with $w_t$ a weight based on the distance $\smash{d_c = \|\mathbf{x}_t - \mathbf{x}_c\|}$ between the center of the triangle $t$ and the center of the fluid cell under consideration (see Figure \ref{fig:PF_1}), 
calculated using the Wendland radial basis function (RBF)~\cite{Wendland1995}
\begin{equation}
    w_t = 
\begin{cases}
\left(1 + 4\dfrac{d_c}{\delta}\right)\left(1-\dfrac{d_c}{\delta}\right)^4 &  \mathrm{if} \ d_c\leq \delta, \\
0  &  \mathrm{if} \ d_c>\delta,
\label{eq:Wendland}
\end{cases}
\end{equation}
where $\delta$ is the radius of the RBF's support. In this work, the 3-ring neighbourhood of triangles around the front-vertex closest to the cell center is considered (see Figure \ref{fig:PF_1}) and a support radius $\delta = 2.5 \Delta x$ is used, as in~\cite{Han2024}. For each local volume fit, this support radius may have to be iteratively adjusted in order for a minimum amount of triangles to contribute non-zero weights. A number of triangles with non-zero weights ranging from 24 to 48 was found to produce accurate paraboloid fits. 
In Eq.~\eqref{eq:volumefit}, the integration domain $\Gamma_t$ is the projection of triangle $t$ onto the ($\mathbf{t}_1$, $\mathbf{t}_2$) plane, and
\begin{equation}
    g_t(x,y) = j_{t,0} + j_{t,1} x + j_{t,2} y
\end{equation}
is the parametric description of the plane containing the triangle. Solving the minimization problem in Eq.~\eqref{eq:volumefit} results in the set of linear equations
\begin{equation}
    \sum_t w_t\left(\int_{\Gamma_t}\left(\mathcal{P}\left(x, y \right)-g_t\left(x, y\right)\right)\  \mathrm{d} A\right)\left(\int_{\Gamma_t} \varphi\  \mathrm{d} A\right)=0 , \quad \varphi \in \left\{ 1, x, y, x^2, xy, y^2 \right\},
    \label{eq:volumefit2}
\end{equation}
which are solved for the variables $\mathbf{a}$.
Using Green's theorem, the area integrals contained in Eq.~\eqref{eq:volumefit2} are transformed into line integrals over the triangle edges. For the detailed derivation of those, the reader is referred to the work of \citet{Jibben2019}.

From these local parabolic approximations, the volume fractions are computed using the closed-form expressions of the first moments of a polyhedron clipped by a paraboloid derived by \citet{Evrard2023}, and available in the open-source Interface-Reconstruction-Library~\cite{Chiodi2022}. 
These analytical expressions have been obtained following successive applications of the divergence theorem so as to transform the 3D geometric moment integrals into a sum of 1D integrals. They return the volume of the intersection of the fluid cell and the space below the local paraboloid approximation of the interface, $V_{\mathcal{P}}$ (see Figure \ref{fig:PF_3}). The discrete indicator function value for the respective fluid cell can therefore be calculated as
\begin{equation}
    \alpha = \frac{V_{\mathcal{P}}}{V_{C}},
\end{equation}
with $V_{C}$ the volume of the fluid cell. Furthermore, the local mean curvature and normal vector of the paraboloid, respectively the approximate local mean curvature and normal vector of the interface in the fluid cell are calculated analytically. \reviewer{The normal vector and curvature are calculated by integrating over the surface of the paraboloid within the fluid cell. This ensures that these quantities are uniquely defined for each cell, as the integration is performed over the exact geometry of the interface represented in the cell. For further details on the implementation and methodology of this evaluation, the reader is referred to the work of \citet{Evrard2023}.} Since the polyhedron intersection algorithm is machine accurate, the only numerical error made when computing these properties arises from the least-squares fitting of the paraboloid to the surface mesh. 
The indicator function value for the remaining full and empty fluid cells can be determined after the interfacial cells have been determined using well established ray-casting or flood-fill algorithms. 

\subsection{Surface tension}

Classically, the force due to surface tension is computed on the surface mesh using a Frenet-Element method \cite{Tryggvason2011a} (or, similarly, a Frenet-Vertex method), with the force on each triangle $t$ given as 
 \begin{equation}
		\mathbf{F}_{\mathrm{s},t} = \sum_e \sigma_e \left(\mathbf{n}_e \times \mathbf{t}_e \right)  l_e ,
\end{equation}
where $\sigma$ is the surface tension coefficient, $\mathbf{t}_e$ and $\mathbf{n}_e$ are tangential and normal vector of the triangle edge $e$, respectively, and $l_e$ is the length of the triangle edge $e$. This force is then interpolated to the fluid mesh, typically using the Peskin cosine kernel given in Eq.~\eqref{eq:peskin}. While this formulation inherently accounts for a spatially varying surface-tension coefficient, which is not considered in this study, the spreading with a smooth kernel yields a smeared-out representation of the force due to surface tension.

The force due to surface tension is proposed to be modelled by the Continuum Surface Force method \cite{Brackbill1992} as a volumetric source term in the momentum equations \eqref{eq:momentum}. The surface tension source term is, thus, given for mesh cell $P$ as
\begin{equation}
    \mathbf{S}_{\sigma,P} = \sigma \kappa_P \nabla \alpha_P \label{eq:csf} ,
\end{equation}
using the indicator function, $\alpha$, and the interface curvature, $\kappa$, obtained from the parabolic reconstruction.
The numerical representation of the interface, modelled as a region with finite thickness, introduces a challenge in utilizing the interface curvature estimate for the computation of the volumetric surface tension force. Despite achieving geometric precision in estimating curvature with the parabolic interface reconstruction, substantial numerical errors may arise due to the spatial variation of curvature in the direction normal to the interface within its finite thickness. For the accurate determination of surface forces, it is imperative to maintain a constant interface curvature in the direction normal to the interface. This consistency is crucial to ensure that the surface force remains invariant in the normal direction. To obtain a representative curvature value across the entire interface region, a weighted averaging approach is employed to smooth the curvature on the fluid mesh. This method aims to mitigate the impact of spatial variations and provide a more reliable curvature estimate for accurate surface force calculations. The weighted smoothing in neighbouring cells of interfacial cells with respect to the interface normal vector is given as \citep{Denner2014a}
\begin{equation}
    \bar{\kappa}_P = \frac{\sum_{Q} \kappa_Q w_Q}{\sum_{Q} w_Q}.
\end{equation}
The weighting factor $w_Q$ is defined as
\begin{equation}
    w_Q = \frac{\left| \mathbf{s}_Q \cdot \mathbf{n}_Q \right|}{\lVert \mathbf{s}_Q \rVert},
\end{equation}
where $\mathbf{n}_Q$ represents the interface normal vector in interfacial cell $Q$ and $\mathbf{s}_Q$ is the vector connecting the cell center of $P$ with the cell center of $Q$.

In order to account for the surface tension source term in the pressure-velocity coupling, $\mathbf{S}_\sigma$ is incorporated in the momentum-weighted interpolation, Eq.~\eqref{eq:mwi}, following the work of \citet{Denner2014a}, with the flux through face $f$ defined as
\begin{multline}
  F_f = \left\{\overline{\mathbf{u}}_{f} \cdot \mathbf{n}_{f} - \hat{d}_f \left[ \frac{p_Q - p_P}{\Delta x} - \frac{\rho^\ast_f}{2} \left( \frac{\nabla p_P}{\rho_P} + \frac{\nabla p_Q}{\rho_Q} \right) \cdot \mathbf{n}_f \right] 
  + \hat{d}_f \sigma \left[\overline{\kappa}_f \frac{\alpha_Q-\alpha_P}{\Delta x} - \frac{\rho^\ast_f}{2} \left(\kappa_P \frac{\nabla \alpha_P}{\rho_P} + \kappa_Q \frac{\nabla \alpha_Q}{\rho_Q} \right) \cdot \mathbf{n}_f \right] \right. \\ 
  + \hat{d}_f \left[\mathbf{S}_{g,f} \cdot \mathbf{n}_f - \frac{\rho^\ast_f}{2} \left(\frac{\mathbf{S}_{g,P}}{\rho_P} + \frac{\mathbf{S}_{g,Q}}{\rho_Q} \right) \cdot \mathbf{n}_{f} \right]
  + \left. \hat{d}_f \frac{\rho^\ast_f}{\Delta t} \left( \vartheta_f^{(t-\Delta t)} - \overline{\mathbf{u}}_{f}^{(t-\Delta t)} \cdot \mathbf{n}_{f} \right) \right\} A_f.
  \label{eq:mwi_disc}
\end{multline}
This treatment of the surface tension source term yields a force-balanced implementation \cite{Denner2014a,Denner2022b,Denner2022c}.

\subsection{Local roughness smoothing}

The proposed local roughness smoothing follows two steps. The first step consists in detecting local surface roughness by measurements of the distance between a front vertex and its geometric Laplacian, and of the variance of the dihedral angles of adjacent triangle normals surrounding a vertex as an indicator of the local curvature variation \cite{Karni2000, Corsini2007}. This is followed by the application of Algorithm 4 of \citet{Kuprat2001} in areas where the measured roughness is above predefined thresholds.

The geometric Laplacian
\begin{equation}
    \mathcal{GL}(\mathbf{x}_i) = \mathbf{x}_i - \frac{\sum_{j \in \mathcal{N}^1_i} w_{ij} \mathbf{x}_j}{\sum_{j \in \mathcal{N}^1_i} w_{ij}},
\end{equation}
proposed by \citet{Karni2000}, measures the smoothness of the front at vertex $\mathbf{x}_i$. The set $\mathcal{N}^1_i$ here represents the one-ring neighbourhood of vertex $\mathbf{x}_i$, and $w_{ij} = 1/\lVert \mathbf{x}_i - \mathbf{x}_j \rVert$ is the inverse distance between vertices $\mathbf{x}_i$ and $\mathbf{x}_j$. The geometric Laplacian represents the difference between vertex $\mathbf{x}_i$ and its Laplacian smoothing equivalent. Hence, the larger it is, the rougher is the surface around $\mathbf{x}_i$. In this work, the threshold magnitude of the geometric Laplacian, above which local roughness smoothing is triggered, is set between 1.2 and 1.5 of the maximum magnitude of the geometric Laplacian of the initial front mesh. 
\reviewerRevtwo{These values may be adjusted based on the specific requirements of the simulation setup.}

A second surface roughness metric is adopted from the work of \citet{Corsini2007}, in which are addressed issues related to the assessment of distortions by watermarked 3D surface meshes. This roughness metric uses statistics of the dihedral angle of adjacent triangles as an indicator for local curvature variation. The dihedral angle \Editor{between two front triangles} is the angle between their normal vectors. On smooth surfaces, the dihedral angles are close to zero since normals vary slowly over the surface. Therefore, for each edge connecting two triangles, a roughness metric $\mathcal{R}_e$ follows as \cite{Corsini2007}
\begin{equation}
    \mathcal{R}_e = 1 - (\mathbf{n}_1 \cdot \mathbf{n}_2),
\end{equation}
where $\mathbf{n}_1$ and $\mathbf{n}_2$ are the normals to the triangles adjacent to edge $e$. The roughness metric statistics per triangle $t$ then follow as
\begin{equation}
    \mathcal{R}_t = \frac{\sum_{i=1}^{3} G_iV_i}{\sum_{i=1}^{3} V_i},
\end{equation}
where $G_i$ is the mean of the roughness $\mathcal{R}_e$ of all edges connected to the vertex $\mathbf{x}_i$ of triangle $t$, and $V_i$ its variance.  
$\mathcal{R}_t$ is able to measure roughness on the scale of a triangle but fails to meaningfully detect roughness at larger scales~\cite{Corsini2007}. This roughness metric is adapted to a per-vertex roughness given as
\begin{equation}
    \mathcal{R}^l_{\mathbf{x}_{i}} = \frac{1}{|\mathcal{N}^l_i|} \sum_{j \in \mathcal{N}^l_i} A_{t_j} \mathcal{R}_{t_j}.
\end{equation}
Here, $\mathcal{N}^l_i$ corresponds to the ring neighbourhood of vertex $\mathbf{x}_i$ with $l$ indicating how many layers of neighbour rings it contains, $|\mathcal{N}^l_i|$ returns the number of triangles in that set, and $A_t$ is the area of triangle $t$. 
In this work, we consider the one- and two-ring neighbourhoods of each vertex, i.e., $l=\{1,2\}$, so as to detect undulations in the close vicinity of that vertex. 
The final roughness metric per vertex is taken as
\begin{equation}
    \mathcal{R}_{\mathbf{x}_{i}} = \max\left(\mathcal{R}^1_{\mathbf{x}_{i}},\mathcal{R}^2_{\mathbf{x}_{i}}\right) .
\end{equation}
As for the geometric Laplacian metric, local roughness smoothing is initiated when $\mathcal{R}_{\mathbf{x}_{i}}$ exceeds 1.2 to 1.5 times the maximum roughness of the initial front mesh.
\reviewerRevtwo{However, the selection of these threshold values should be guided by the characteristics of the test case, as the appropriate choice depends on factors such as mesh resolution, the scale of the features to be smoothed, and the desired level of smoothing.}
\Editor{For further clarity, in \ref{AppendixA}, we have included a detailed derivation on how to relate the curvature to the roughness metric. This derivation can serve as a guideline for selecting an appropriate threshold parameter in the smoothing procedure.}

The local roughness smoothing is adopted from the work of \citet{Kuprat2001}. In their Algorithm 4, they propose a volume-conserving smoothing algorithm that relaxes the vertices of an edge on surface meshes. In our work we use this algorithm for each front mesh edge whose vertices exceed the thresholds of the two previously explained roughness metrics. In the following the most important steps of the smoothing algorithm are explained. For more detailed derivations the reader is referred to \citep{Kuprat2001}.

Consider an edge $e$ with its two vertices $\mathbf{c}_i, i\in\{1,2\}$. The new positions of the vertices are calculated as
\begin{equation}
    \mathbf{x}_{i,\mathrm{new}} = \mathbf{x}_i^s + h \hat{\mathbf{n}},
\end{equation}
with $\mathbf{x}_i^s$ obtained by simultaneous Laplacian smoothing of both edge vertices,
\begin{align}
    \mathbf{x}_1^s & = \frac{1}{|\mathcal{N}^1_1|} \left( \mathbf{x}_2^s + \sum_{j \in \mathcal{N}^1_1 \backslash \mathbf{x}_2 } \mathbf{x}_j \right)
    \label{Eq:LapSmooth1}\\
    \mathbf{x}_2^s & = \frac{1}{|\mathcal{N}^1_2|} \left( \mathbf{x}_1^s + \sum_{j \in \mathcal{N}^1_2 \backslash \mathbf{x}_1} \mathbf{x}_j \right).
    \label{Eq:LapSmooth2}
\end{align}
Substituting Eq.~\eqref{Eq:LapSmooth2} into Eq.~\eqref{Eq:LapSmooth1} yields
\begin{equation}
    \mathbf{x}_1^s = \frac{1}{|\mathcal{N}^1_1||\mathcal{N}^1_2| - 1} \sum_{j \in \mathcal{N}^1_2 \backslash \mathbf{x}_1} \mathbf{x}_j + \frac{|\mathcal{N}^1_2|}{|\mathcal{N}^1_1||\mathcal{N}^1_2| - 1} \sum_{j \in \mathcal{N}^1_1 \backslash \mathbf{x}_2 } \mathbf{x}_j,
    \label{eq:LapSmooth3}
\end{equation}
and $\mathbf{x}_2^s$ can be obtained by substituting the result of Eq.~\eqref{eq:LapSmooth3} into Eq.~\eqref{Eq:LapSmooth2}. The factor $h$ and direction $\hat{\mathbf{n}}$ are chosen such that volume is conserved and $\lVert h \hat{\mathbf{n}} \rVert$ is minimal. Following \citet{Kuprat2001}, $\hat{\mathbf{n}}$ is derived as
\begin{equation}
    \hat{\mathbf{n}} = \frac{\mathbf{A}_1 + \mathbf{A}_2 + \mathbf{y} \times ((\mathbf{x}_1^s - \mathbf{x}_1) - (\mathbf{x}_2^s - \mathbf{x}_2))}{\lVert \mathbf{A}_1 + \mathbf{A}_2 + \mathbf{y} \times ((\mathbf{x}_1^s - \mathbf{x}_1) - (\mathbf{x}_2^s - \mathbf{x}_2)) \rVert}
\end{equation}
and $\lVert h \hat{\mathbf{n}} \rVert$ is minimal when
\begin{equation}
  h = - \frac{(\mathbf{x}_1^s - \mathbf{x}_1) \cdot \mathbf{A}_1 + (\mathbf{x}_2^s - \mathbf{x}_2) \cdot \mathbf{A}_2 + (\mathbf{x}_1^s - \mathbf{x}_1) \cdot \left( \mathbf{y} \times (\mathbf{x}_1^s - \mathbf{x}_1) \right) }{\hat{\mathbf{n}} \cdot (\mathbf{A}_1 + \mathbf{A}_2 + \mathbf{y} \times ((\mathbf{x}_1^s - \mathbf{x}_1) - (\mathbf{x}_2^s - \mathbf{x}_2)))},  
\end{equation}
with 
\begin{equation}
    \mathbf{A}_i = \sum_{j \in \mathcal{N}^1_i} \mathbf{z}_i^{j} \times \mathbf{z}_i^{j + 1}
\end{equation}
being the sum of the cross products of the edges $\mathbf{z}$ connected to vertex $i$ and $\mathbf{y}$ is the vector connecting the two remaining vertices of the triangles connected to the edge under consideration for the smoothing algorithm. In summary, the algorithm of \citet{Kuprat2001} can be explained as an extension of classic Laplacian smoothing combined with a TSUR-3D-like method to ensure local volume conservation.

The proposed two-step local surface roughness smoothing method eliminates the need for pre-determined time intervals at which undulation removal or roughness smoothing algorithms are utilized. This is achieved by making small, localized adjustments to the front without changing its overall shape. It also prevents mesh folding and improves the quality of the front mesh, allowing accurate computations of geometric properties.

\section{Results}
\label{sec:results}
In this section, the results of various test cases are presented to verify, validate, and demonstrate the capabilities of the proposed sharp front-tracking method for interfacial flows.

\subsection{Stationary droplet}
The accurate computation of the interface properties and \reviewer{the force balancing} of the proposed sharp FT method with respect to the surface tension force is validated and compared to the classic FT method for a stationary spherical interface. In order to isolate the force-balancing methodology and the computation of the geometrical properties of the interface, certain test case conditions have to be ensured: other body forces, as for instance gravity, should be absent and the interface should remain in a mechanical equilibrium. For interfacial flows, the most common test case for this scenario is the Laplace equilibrium~\cite{Popinet2018}. Under the absence of any other body forces except the surface tension force, a spherical interface is surrounded by a quiescent fluid. Under exact physical conditions, the interface would remain stationary and only the pressure jump across the interface, given by the Young-Laplace equation
\begin{equation}
    \Delta p_{\text{exact}} = \sigma \kappa
\end{equation}
would develop. Therefore, errors in the pressure jump across the interface and velocity magnitudes larger than the employed solver tolerance or machine precision are the result of a numerical imbalance between pressure gradient and the surface tension force and arise from numerical inaccuracies in curvature computation or the velocity interpolation to the Lagrangian vertices. Within the parabolic interface reconstruction, numerical inaccuracies only arise from numerical errors in the computation of the least-squares paraboloid fit of the surface mesh and the resulting deviation from the exact mean curvature, which for a sphere is
\begin{equation}
    \kappa_{\text{exact}} = \frac{2}{r},
\end{equation}
with $r$ being the initial radius of the sphere. Velocity magnitudes larger than the solver tolerance, arising from the numerical inaccuracies, are commonly referred to as parasitic or spurious currents and should be dissipated over time by viscosity. Therefore, the most important parameter for the Laplace equilibrium test case are the Laplace number 
\begin{equation}
    \mathrm{La} = \frac{d \sigma \rho_f}{\mu_f^2}
\end{equation}
and the viscous time scale
\begin{equation}
    \tau_{\mu} = \frac{\rho_f d^2}{\mu_f},
\end{equation}
describing the ratio of surface tension to viscous forces and the viscous dissipation time scale, where subscript $f$ refers to the fluid surrounding the droplet. The evolution of the parasitic currents can be described and non-dimensionalized by the capillary number
\begin{equation}
    \mathrm{Ca} = \frac{\mu_f |\mathbf{u}_{\mathrm{ref}}|}{\sigma},
\end{equation}
describing the ratio of viscous drag forces to surface tension forces. In the remainder of this work we use the capillary numbers based on the root-mean-square (rms) velocity $\mathrm{Ca}_{\mathrm{rms}}$ and the maximum velocity $\mathrm{Ca}_{\mathrm{max}}$.
\reviewerRevtwo{The rms velocity is defined as
\begin{equation}
    u_{\mathrm{rms}} = \sqrt{\frac{1}{N} \sum_{i=1}^{N} |\mathbf{u}_i|^2},
\end{equation}
where $N$ is the total number of mesh cells, and $\mathbf{u}_i = (u_i, v_i, w_i)$ represents the velocity components at each mesh cell and the maximum velocity is computed as
\begin{equation}
    u_{\max} = \max_{i \in {1, \dots, N}} |\mathbf{u}_i|.
\end{equation}
}

The setup of this test case is adapted from \cite{Popinet2009, Abadie2015}, with the difference that in this study no symmetry is employed and the complete 3D interface is simulated. The cubic domain with an equidistant Cartesian fluid mesh has the size $2.0 \times 2.0 \times 2.0$ and the initial sphere has a diameter of $0.8$, centered at $[1.0 \ 1.0 \ 1.0]$. The density for both fluids and the surface tension coefficient are set to unity, and the dynamic viscosity for both fluids is based on the Laplace number. The Laplace numbers considered for this test case are $\mathrm{La} = \{ 120, 1200, 12000, 48000, 120000 \}$. The time step for all test cases is based on the capillary time step constraint \cite{Denner2015}.

\begin{figure}
    \centering
    \includegraphics[scale=0.9]{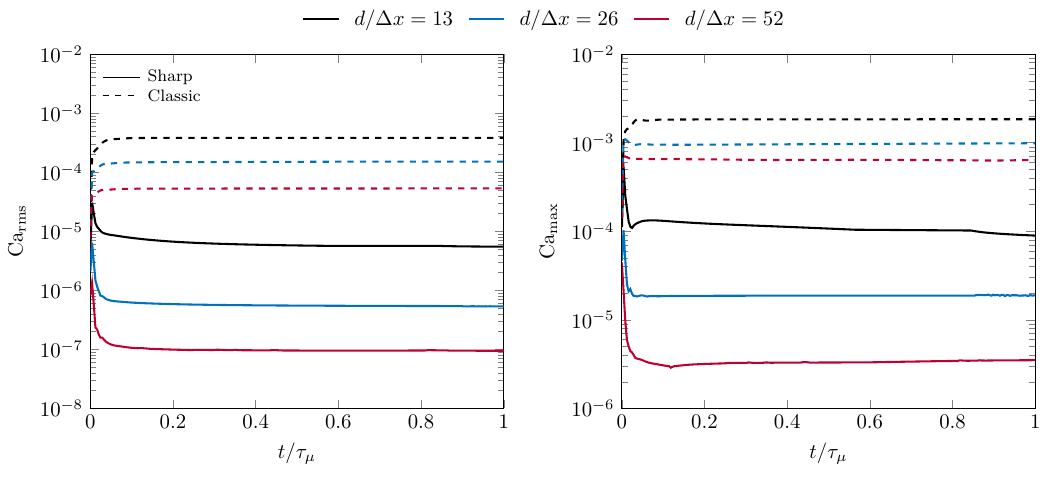}
    \caption{Evolution of the capillary number of the 3D Laplace equilibrium with Laplace number La = 120 for the proposed sharp front tracking and classic front tracking for different fluid mesh spacings. The ratio of fluid mesh spacing to front mesh spacing $\Delta x/l_e$ is kept constant at 1. The capillary number is based on the root-mean-square (rms) of the parasitic velocities in the left plot and based on the maximum velocity in the right plot. The time is normalized by the viscous timescale $\tau_\mu = \rho d^2 / \mu$.}
    \label{fig:StationaryDroplet_La120_FluidMeshConvergence}
\end{figure}

\begin{figure}
    \centering
    \includegraphics[scale=0.9]{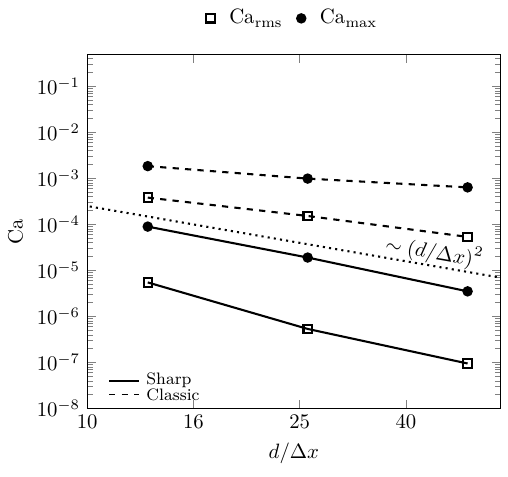}
    \caption{Logarithmic plot comparing the capillary numbers at $t/\tau_{\mu} = 1$ of the 3D Laplace equilibrium with Laplace number La = 120 for the proposed sharp front tracking and classic front tracking for different fluid mesh spacings. The dotted line indicates second-order convergence.}
    \label{fig:StationaryDroplet_La120_FluidMeshConvergence_LogLog}
\end{figure}

Figure \ref{fig:StationaryDroplet_La120_FluidMeshConvergence} shows the evolution of the capillary numbers based on the rms-velocity (left plot) and the maximum velocity (right plot) over the viscous time scale for a Laplace number of $120$. The classic FT method and the proposed sharp FT method are compared for three different fluid mesh resolutions. The ratio of fluid cell spacing to the average triangle edge length of the front mesh, $\Delta x/l_e$, is kept constant at $1$ for all cases in this figure. The plots show that both capillary numbers increase and then reach a steady state with classic front tracking, whereas for the sharp front tracking, the capillary numbers first increase and then decrease until a steady state is reached. As shown in Figure \ref{fig:StationaryDroplet_La120_FluidMeshConvergence_LogLog}, both capillary numbers for the sharp front tracking converge with second order convergence for increasing resolution, whereas the convergence rate for the classic front tracking is second order at best. Overall, the capillary number based on the rms-velocity for the proposed sharp front tracking is approximately two orders of magnitudes smaller compared to classic front tracking for all three resolutions and at least one order of magnitude for the capillary number based on the maximum velocity.


\begin{figure}
    \centering
    \includegraphics[scale=0.9]{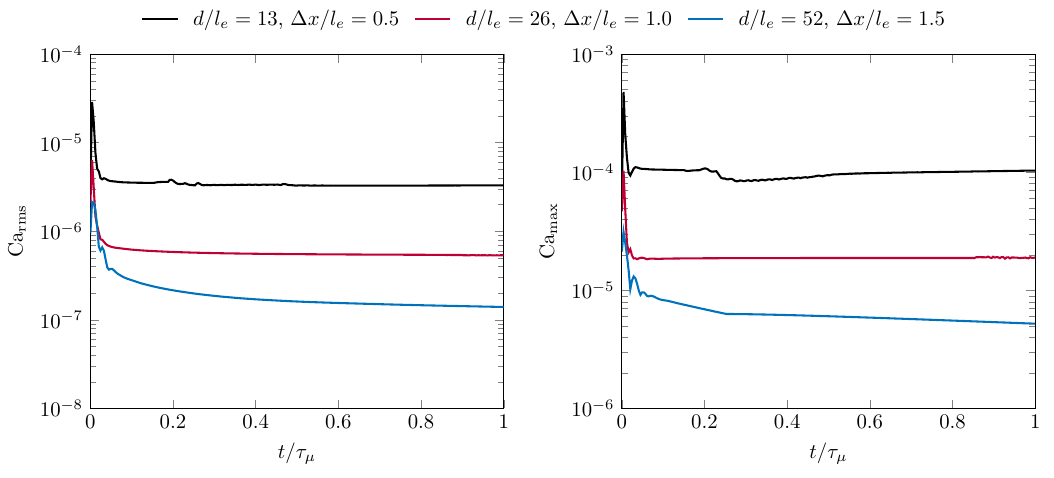}
    \caption{Evolution of the capillary number of the 3D Laplace equilibrium with Laplace number La = 120 for the proposed sharp front tracking for different front mesh spacings, given in ratios of the initial diameter $d$ or the fluid mesh spacing $\Delta x$ to the average triangle edge length $l_e$. The number of fluid mesh cells per droplet diameter is kept constant at $d/\Delta x = 26$. The capillary number is based on the root-mean-square (rms) of the parasitic velocities in the left plot and based on the maximum velocity in the right plot. The time is normalized by the viscous timescale $\tau_\mu = \rho d^2 / \mu$.}
    \label{fig:StationaryDroplet_La120_FrontMeshConvergence}
\end{figure}

Figure \ref{fig:StationaryDroplet_La120_FrontMeshConvergence} shows the evolution of both capillary numbers for the proposed sharp FT method at a Laplace number of $120$ for different front mesh resolutions, but for a constant fluid mesh resolution of $d/\Delta x = 26$. The front mesh resolution is given in diameter to average triangle edge length ratio $d/l_e$ and in the ratio of fluid mesh resolution to average triangle edge length $\Delta x/l_e$. Both capillary numbers decrease with second order for increasing front mesh resolution. Figures \ref{fig:StationaryDroplet_La120_FluidMeshConvergence} and \ref{fig:StationaryDroplet_La120_FrontMeshConvergence} show that the proposed sharp FT method converges well with increasing fluid mesh and front mesh resolution. The reasons are the reduced numerical errors in the fluid velocity interpolation due to a higher resolved fluid mesh and the reduced numerical errors in the computation of the curvature based on the parabolic fit due to higher resolved front mesh. Nonetheless, it should be mentioned here that a $\Delta x/l_e$-ratio larger than $1$ may only be used with caution. If the front mesh is finer than the fluid mesh, the Lagrangian vertex motion may not be strongly correlated to any force and the resulting velocity on the fluid mesh, which may give rise to surface undulations and roughness on the front mesh.  

\begin{figure}
    \centering
    \includegraphics[scale=0.9]{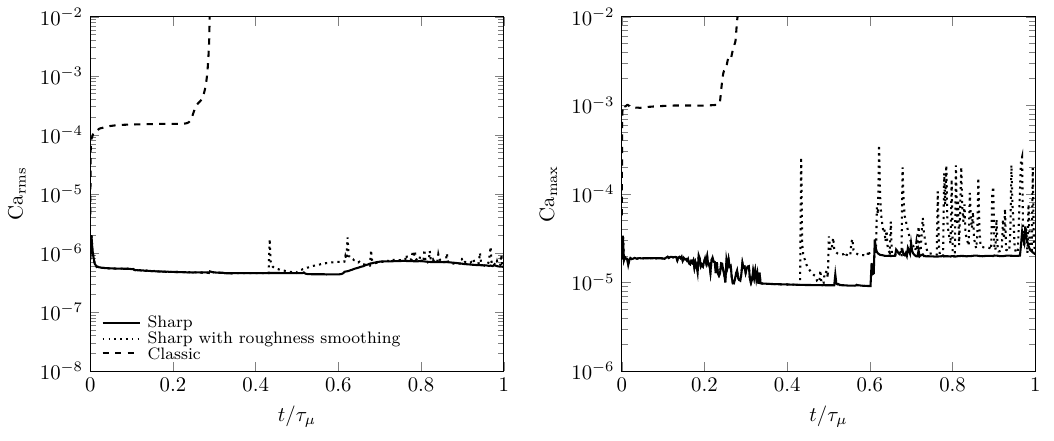}
    \caption{Evolution of the capillary number of the 3D Laplace equilibrium with Laplace number La = 1200 for the proposed sharp front tracking and classic front tracking for the ratios $d/\Delta x = 26$ and $\Delta x/l_e = 1$. The capillary number is based on the root-mean-square (rms) of the parasitic velocities in the left plot and based on the maximum velocity in the right plot. The time is normalized by the viscous timescale $\tau_\mu = \rho d^2 / \mu$.}
    \label{fig:StationaryDroplet_La1200.pdf}
\end{figure}

For all Laplace numbers above $120$, only the mesh resolutions of $d/\Delta x =26$ and $\Delta x/l_e =1$ are considered.
In Figure \ref{fig:StationaryDroplet_La1200.pdf}, the evolution of the capillary numbers for the classic and sharp FT methods is shown at a Laplace number of $1200$. The classic FT method is not able to produce a stable result and, after a short time, the parasitic currents increase rapidly without bounds. The sharp FT method is, however, still able to produce a stable result and the interface remains spherical throughout the simulation. The capillary number based on the maximum velocity starts to oscillate and a jump for both capillary numbers is observed at $t/\tau_\mu \approx 0.6$. The reason for this is that the spherical interface slowly starts to move in space around the center position, as a result of the non-symmetric nature of the front mesh and numerical errors in the velocity interpolation. In addition, small imbalances and jumps in the parasitic currents may cause a switch of the closest Lagrangian vertex with respect to a cell center of the fluid mesh for the computation of the parabolic fit. Even though the current least-squares weighting function based on the distance between the point under consideration and the cell center of the fluid mesh mitigates this phenomenon, a change in the closest vertex can also change the neighbourhood information that is handed to the parabolic fit and, therefore, may lead to a slightly different curvature. This, in turn, may lead to a sudden small force imbalance. Whether a change in neighbourhood information occurs, depends on the order of the direct neighbourhood and the preset minimum number of triangles within the weighting distance. A minimum number of triangles within the weighting distance that is close to the total number of triangles in the chosen order of neighbourhood can lead to a change of neighbourhood information, yet also leads to a more accurate paraboloid fit. In contrast, a smaller minimum number of triangles within the weighting distance reduces the possibility of a change of neighbourhood information, but may lead to a slightly less accurate paraboloid fit. Overall, such small imbalances can give rise to a small increase in parasitic currents and the system needs to rebalance itself again. A similar behaviour can be observed for the sharp front tracking with roughness smoothing, shown as the dotted line in Figure  \ref{fig:StationaryDroplet_La1200.pdf}. Every time instance the roughness smoothing is triggered and the Lagrangian vertices undergo a smoothing operation, the parasitic currents increase due to a resulting force imbalance originating from a sudden change in vertex position, and the entire system needs to find its balance again. The comparison between sharp front tracking with and without roughness smoothing shows that for a Laplace number of $1200$, the sharp front tracking is stable enough without roughness smoothing, since enough kinetic energy of the parasitic currents is dissipated by viscosity.        

\begin{figure}
    \centering
    \includegraphics[scale=0.9]{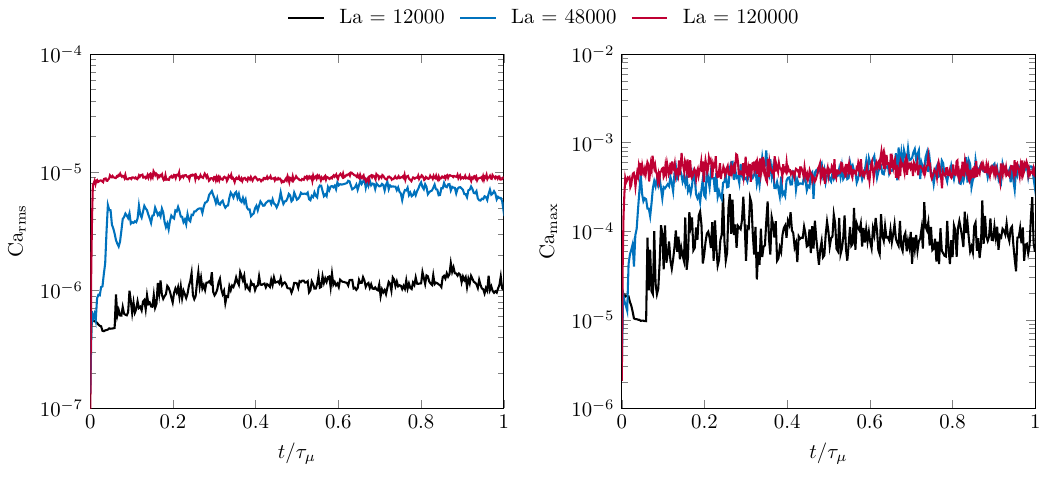}
    \caption{Evolution of the capillary number of the 3D Laplace equilibrium with Laplace number $\text{La} = 12000-120000$ for the proposed sharp front tracking for the ratios $d/\Delta x = 26$ and $\Delta x/l_e = 1$. The capillary number is based on the root-mean-square (rms) of the parasitic velocities in the left plot and based on the maximum velocity in the right plot. The time is normalized by the viscous timescale $\tau_\mu = \rho d^2 / \mu$. Local roughness smoothing is triggered and operating throughout the simulations. Classic front tracking is diverging rapidly and, hence, not shown.}
    \label{fig:StationaryDroplet_La12000_120000.pdf}
\end{figure}

The Laplace numbers $12000$, $48000$ and $120000$ are shown for the proposed sharp FT method in Figure \ref{fig:StationaryDroplet_La12000_120000.pdf}. The classic FT method diverges shortly after the start of the simulation and is not able to attain a steady state, therefore it is not shown for these Laplace numbers. For the sharp front tracking, the local roughness smoothing is operating throughout the simulation to keep the interface spherical for all three Laplace numbers. Without the local roughness smoothing, the sharp front tracking diverges as well. It can be seen that with increasing Laplace number, the magnitudes of both capillary numbers increases, whereas the amplitudes of the oscillations around the quasi steady-state decrease.

\begin{figure}
    \centering
    \includegraphics[scale=0.9]{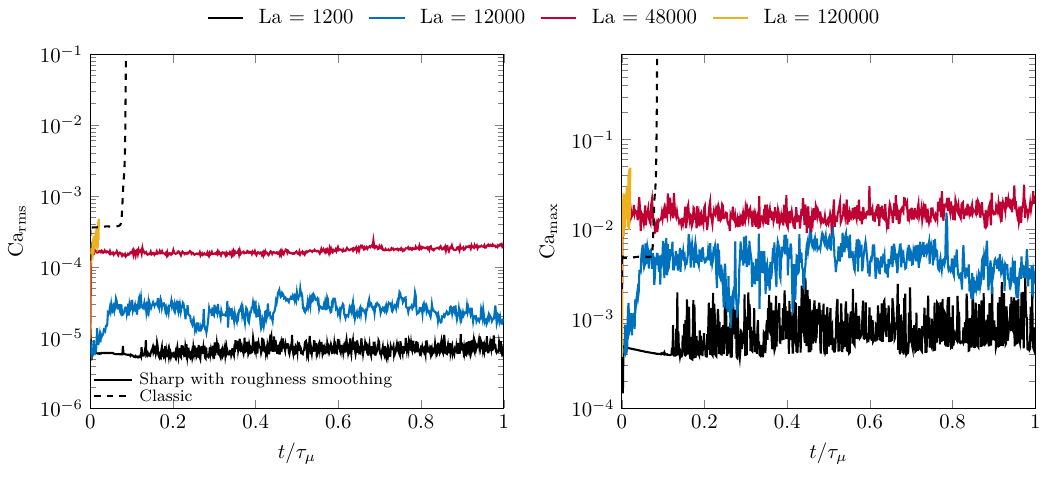}
    \caption{\reviewer{Evolution of the capillary number of the 3D Laplace equilibrium with Laplace number $\text{La} = 1200-120000$ for the proposed sharp front tracking for ratios $d/\Delta x = 26$ and $\Delta x/l_e = 1$. The density ratio is $ \rho_f / \rho_d =1000$ and the viscosity ratio is $\mu_f / \mu_d = 31.6$, ensuring that the Laplace numbers for the fluid and the droplet are equal. The capillary number is based on the root-mean-square (rms) of the parasitic velocities in the left plot and based on the maximum velocity in the right plot. The time is normalized by the viscous timescale $\tau_\mu = \rho d^2 / \mu$. Local roughness smoothing is triggered and operating throughout the simulations. The simulations using classic front tracking diverge rapidly and, hence, are only shown for $\mathrm{La} = 1200$.}}
    \label{fig:StationaryDroplet_Density_1000.pdf}
\end{figure}

\reviewer{In the literature, realistic density ratios on the order of $\mathcal{O}(10^3)$, which are representative of most real-world applications, have not been extensively studied with front-tracking methods. However, we recognize the importance of demonstrating the robustness of our approach under more realistic conditions. To address this, Figure \ref{fig:StationaryDroplet_Density_1000.pdf} shows the results of the evolution of the capillary number for Laplace numbers from 1200 to 120000 for a density ratio of $\mathcal{O}(10^3)$, showcasing the capability of the proposed front-tracking method to handle more challenging scenarios compared to the classic front-tracking method. Classic front tracking is not able to handle such density ratios, whereas the proposed method is able to achieve stable results throughout, except for the Laplace number of 120000. In comparison to a unit density ratio, the magnitude of the capillary number is higher for larger density ratios and roughness smoothing is required throughout.}

For the Laplace equilibrium test case of a stationary droplet it can be concluded that the proposed sharp FT method is able to retain a force balance between surface tension and the associated pressure jump, and yields a stable solution for a wide range of Laplace numbers, up to at least a Laplace number of $120000$. Likewise, the magnitude of the parasitic currents produced by the sharp FT method are orders of magnitude smaller than those produced by the classic FT method for cases where classic FT can converge at all. Due to the sharp interface representation on the fluid mesh, the sharp FT method is generally more sensitive to distortions. However, the proposed local roughness smoothing addresses this issue, providing a solution and expanding the range of parameters that can be effectively simulated.

\subsection{Translating droplet}
\begin{figure}
	\centering
	\subcaptionbox{$\mathrm{La} = 120$\label{fig:Translating_Droplet_La120}}{\includegraphics[width=11.5cm]{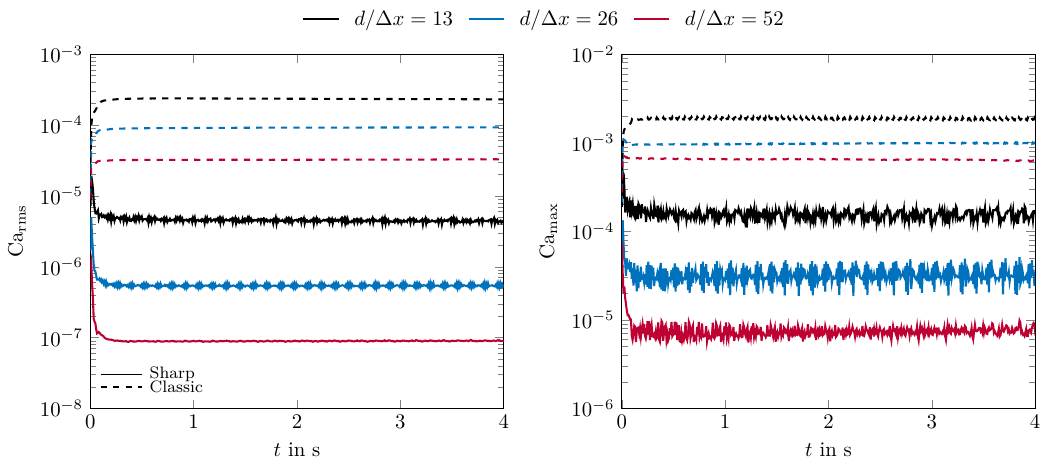}}\\
	\subcaptionbox{$\mathrm{La} = 1200$\label{fig:Translating_Droplet_La1200}}{\includegraphics[width=11.5cm]{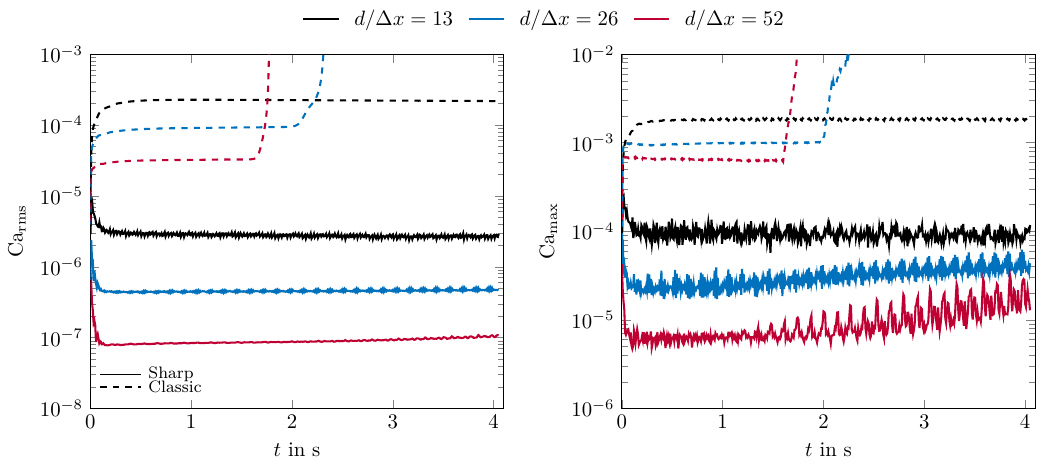}}\\
 \subcaptionbox{$\mathrm{La} = 12000$\label{fig:Translating_Droplet_La12000}}{\includegraphics[width=11.5cm]{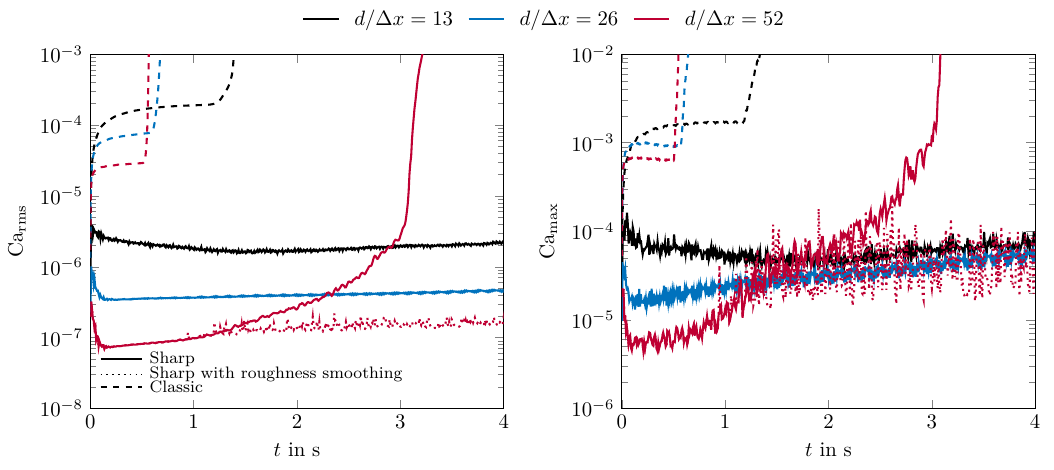}}\\
 \subcaptionbox{$\mathrm{La} = 120000$\label{fig:Translating_Droplet_La120000}}{\includegraphics[width=11.5cm]{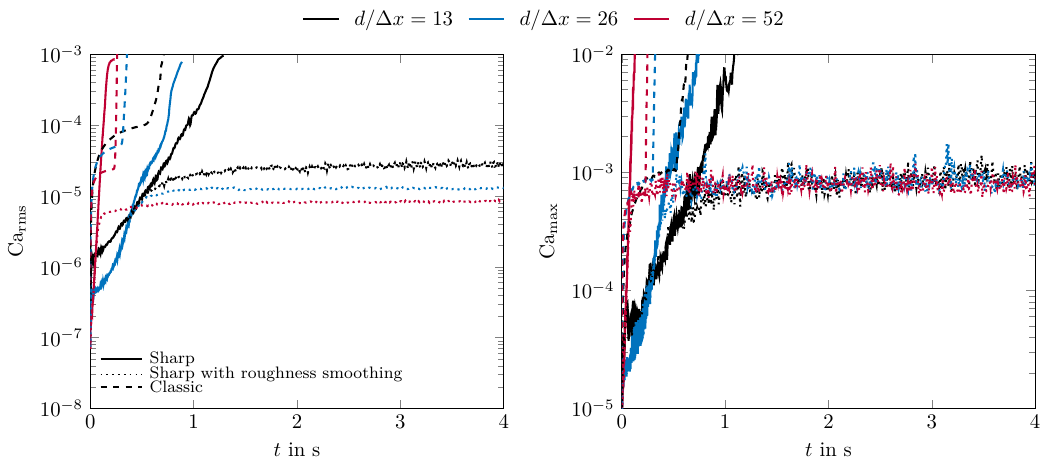}}\\
	\caption{\reviewer{Evolution of the capillary numbers of the translating droplet testcase for Laplace numbers $\mathrm{La}=\{120,1200,12000,120000\}$ for the proposed sharp front tracking and classic front tracking for different fluid mesh spacings. The ratio of fluid mesh spacing to front mesh spacing is $\Delta x/l_e=1$. The capillary numbers in the left plots are based on the root-mean-square (rms) of the spurious flow velocities in the reference frame of the droplet, whereas the capillary numbers in the right plots are based on the maximum velocity in the reference frame of the droplet. The droplets are translated for 10 droplet diameters.}}
	\label{fig:TranslatingDroplets}
\end{figure}

\begin{figure}
    \centering
    \includegraphics[scale=0.9]{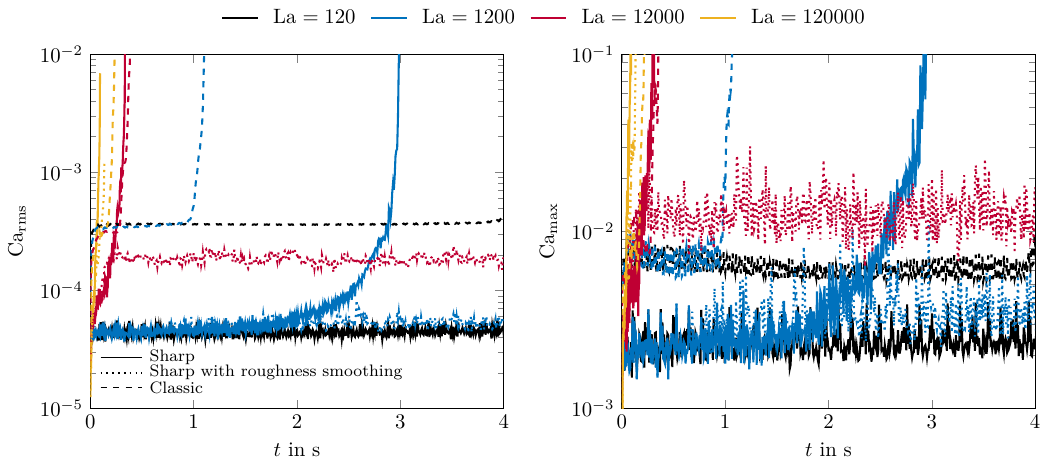}
    \caption{\reviewer{Evolution of the capillary numbers of the translating droplet testcase for Laplace numbers $\mathrm{La}=\{120,1200,12000,120000\}$ for the proposed sharp front tracking and classic front tracking for the ratios $d/\Delta x = 13$ and $\Delta x/l_e = 1$. The density ratio is $ \rho_f / \rho_d =1000$ and the viscosity ratio is $\mu_f / \mu_d = 31.6$, ensuring that the Laplace numbers for the surrounding fluid and the droplet are equal. The capillary numbers in the left plots are based on the root-mean-square (rms) of the spurious flow velocities in the reference frame of the droplet, whereas the capillary numbers in the right plots are based on the maximum velocity in the reference frame of the droplet. The droplets are translated for 10 droplet diameters.}}
    \label{fig:TranslatingDroplet_Density_1000.pdf}
\end{figure}

\reviewer{
In order to reflect a more practical scenario and test the combined accuracy of front advection and surface tension computation, the Laplace equilibrium for the stationary droplet is now adapted to a uniformly translated spherical interface. Theoretically, the exact Laplace equilibrium solution in the moving frame of reference for the translating droplet is unchanged as compared to the stationary droplet, but additional numerical errors in the front advection  induce fluctuating errors in the curvature computation, which in turn will influence the velocity field and the parasitic currents \cite{Popinet2009}. For this test case, the majority of the setup, the initial and boundary conditions are adopted from the work of \citet{Tolle2020}. The droplet has an initial diameter of $d=0.4$ and its center is located at $[0.5 \ 0.5 \ 0.4]$ within a cubic domain of size $2.5d \times 2.5d \times 12.0d$. The constant uniform velocity on the equidistant Cartesian fluid mesh is set to $\mathbf{u} = [0.0 \ 0.0 \ 1.0]$ and the droplet is advected by ten droplet diameters. The material properties and non-dimensional numbers are the same as for the stationary droplet, the only difference is that the background velocity is subtracted from the reference velocities of the capillary numbers. The capillary numbers represent the parasitic currents in the moving frame of reference for the translating droplet and, hence, represent the velocity deviations from the prescribed uniform velocity field.

Figure \ref{fig:TranslatingDroplets} shows the evolution of the capillary numbers over time for the classic and sharp front tracking methods for different fluid mesh resolutions at a constant front mesh resolution of $\Delta x/l_e = 1$, for the Laplace numbers $\mathrm{La}=\{120, 1200, 12000, 120000\}$. The proposed sharp front tracking method shows a second-order convergence rate for $\mathrm{Ca}_{\mathrm{rms}}$ for increasing mesh resolution and approximately second-order convergence for $\mathrm{Ca}_{\mathrm{max}}$ up to Laplace numbers of $\text{La} = 1200$, without the need of any roughness smoothing. The magnitudes of the capillary numbers for the sharp front tracking are, similar to the stationary droplet, approximately to two orders of magnitudes smaller compared to classic front tracking. Classic front tracking leads to stable results only for $\text{La} = 120$ and fails to converge for $\text{La} \geq 1200$ . At $\text{La} = 12000$, the sharp front tracking starts to become unstable for the highest resolution, but sharp front tracking together with roughness smoothing yields a converged result. For $\text{La} = 120000$, only sharp front tracking with roughness smoothing leads to a stable solution, even though the capillary numbers do not further decrease  with an increasing fluid mesh resolution. 

Figure \ref{fig:TranslatingDroplet_Density_1000.pdf} shows the evolution of the capillary numbers over time for a density ratio of $\mathcal{O}(10^3)$ and $\mathrm{La}=\{120,1200,12000,120000\}$. Similar observations as for the stationary Laplace equilibrium with a density ratio of $\mathcal{O}(10^3)$ can be made for the translating droplet test case. The classic front-tracking method is only able to achieve a stable result for $\text{La} = 120$ and fails for all higher Laplace numbers, whereas the proposed method with roughness smoothing is able to achieve stable results for all Laplace numbers, except for the highest Laplace number of 120000.  

In the work of \citet{Abadie2015}, various VOF and level-set methods are compared for the translating droplet test case in 2D with $\text{La} = 12000$ and $d/\Delta x = 32$, and \citet{Tolle2020} provide results for their hybrid level-set/front-tracking method for the same parameters in 3D. \citet{Abadie2015} show that $\mathrm{Ca}_{\mathrm{max}}$ is in the range of $\mathcal{O}(10^{-4}) < \mathrm{Ca}_{\mathrm{max}} < \mathcal{O}(10^{-3})$ for the considered VOF methods and in the range of $\mathcal{O}(10^{-6}) < \mathrm{Ca}_{\mathrm{max}} < \mathcal{O}(10^{-5})$ for the considered level-set methods. The hybrid method of \citet{Tolle2020}, however, achieves $\mathrm{Ca}_{\mathrm{max}} \approx 2.4\cdot10^{-5}$. With a comparable but slightly coarser resolution of $d/\Delta x = 26$, the proposed sharp front-tracking method achieves $\mathrm{Ca}_{\mathrm{max}} \approx 2.4\cdot10^{-5}$, similar to the 
hybrid method of \citet{Tolle2020}, in the range of the level-set methods  and better than the VOF methods considered by \citet{Abadie2015}, without the need of a hybrid method. Nonetheless, it is worth mentioning that the 2D geometric VOF with PPIC of \citet{Remmerswaal2022} demonstrates the ability to decrease the parasitic currents down to machine accuracy for the stationary as well as for the translating droplet test case, which is not possible with the proposed sharp front-tracking method.   
}

\subsection{Oscillating droplet}
In the previous test cases, the interfaces remains spherical. The next natural step is to test the proposed sharp FT method also for interface deformation and a correct surface-tension-driven exchange of potential and kinetic energy, for which an oscillating drop is a common test case. Neglecting the influence of gravity, a droplet is initialized as a prolate ellipsoid and, because of the surface tension, starts to oscillate until it reaches spherical equilibrium. \citet{Lamb1932} provides analytical solutions for small-amplitude oscillations of droplets, where the oscillation frequency of an inviscid droplet is given as
\begin{equation}
    \omega_n = \sqrt{\frac{n(n+1)(n-1)(n+2)\sigma}{[(n+1)\rho_{\mathrm{d}} + n \rho_{\mathrm{a}}]R_0^3}},
\end{equation}
with $n$ being the mode number, $R_0$ the unperturbed radius of the droplet, and the subscripts ${\mathrm{d}}$ and ${\mathrm{a}}$ correspond to the droplet and ambient fluid, respectively. The oscillation amplitude
\begin{equation}
    a_n (t) = a_0 e^{-\gamma t}, \,\,\, \gamma = \frac{(n-1)(2n+1)\mu_{\mathrm{d}}}{R_0^2 \rho_{\mathrm{d}}}
\end{equation}
decreases over time and the initial droplet shape can be defined as
\begin{equation}
    R(\theta,t)=R_0+\epsilon P_n(\cos{\theta})\sin{(\omega_n t)}, \,\,\, \theta \in [0, 2\pi],
\end{equation}
where $P_n$ denotes the $n$-th order Legendre polynomial. The considered droplet has an unperturbed radius of $R_0=1$ and is centered at $[2.0 \ 2.0 \ 2.0]$ within a cubic domain of size $4.0 \times 4.0 \times 4.0$. Furthermore, $n=2$, $\epsilon = 0.025$, $t=\pi/(2\omega_n)$ and the material properties of the fluids are $\rho_{\mathrm{d}} = 10$, $\rho_{\mathrm{a}} = 0.1$, $\mu_{\mathrm{d}} = \{0.5, 0.05 \}$, $\mu_{\mathrm{a}} = 0.0005$ and $\sigma = 10$. The time step is again based on the capillary time step constraint \cite{Denner2015}. The boundaries of the cubic domain are set to no-slip walls and the initial velocity and pressure field are set to $0$. The temporal evolution of the radius in the $x$-direction is measured as half the distance between the vertices of the front mesh with the minimum and maximum position value in the $x$-direction.

\begin{figure}
    \centering
	\subcaptionbox{$\mu_d = 0.5$\label{fig:mu_d_05}}{\includegraphics[width=8cm]{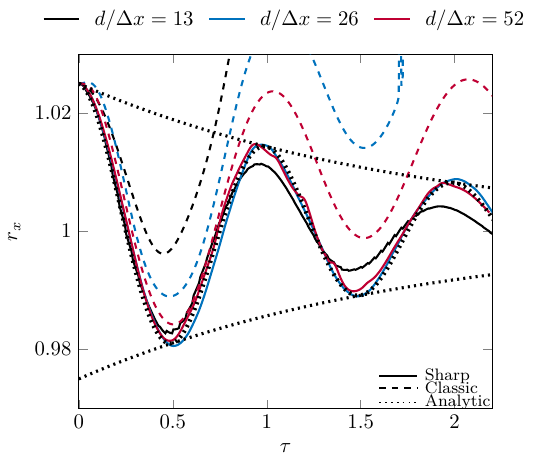}}
	\subcaptionbox{$\mu_d = 0.05$\label{fig:mu_d_005}}{\includegraphics[width=8cm]{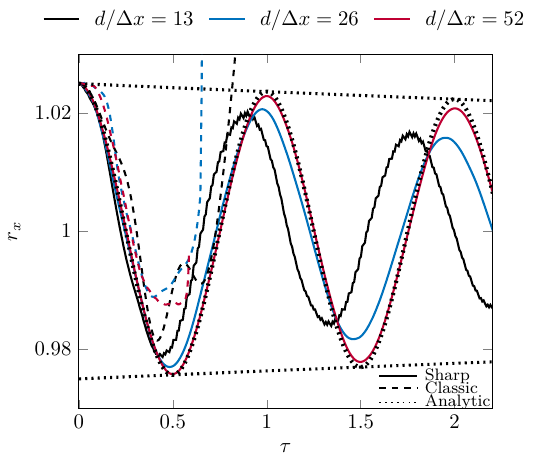}}\\
    \caption{Temporal evolution of the radius in x-direction $r_x$ for the proposed sharp front tracking and classic front tracking for the oscillating droplet testcase for two different droplet viscosities. Time is normalized with the analytical period $2\pi / \omega_2$. The dotted lines represent the analytic solution of the droplet radius evolution and the exact decaying envelope of the amplitude ($R_0 + a_2(t)$ and $R_0 - a_2(t)$).}
    \label{fig:OscillatingDroplets}
\end{figure}

Figure \ref{fig:OscillatingDroplets} shows the temporal evolution of the droplet radius for the two considered droplet viscosities, using the classic FT method and the proposed sharp FT method for different resolutions of the fluid mesh, as well as the analytic reference solution. The resolution of the front mesh is $\Delta x/l_e=1$. It can be seen clearly, that the classic FT method is not able to simulate the oscillation of the drop for the tested resolutions. For the droplet with the larger viscosity, the classic FT method is able to simulate approximately one period of the droplet oscillation on the fluid mesh with the highest resolution, before the simulation diverges due to parasitic currents. The proposed sharp FT method converges towards the analytical solution for both the period and the amplitude with increasing fluid mesh resolution. However, the sharp FT method requires the proposed local roughness smoothing to achieve a stable result for this test case. The roughness smoothing is also the reason for the spurious oscillations in the temporal evolution of the radius, since even very small adjustments in the vertex position have a visible effect on the radius calculation because of small oscillation amplitudes relative to the droplet size. Overall, the sharp FT method leads to accurate results in predicting the droplet oscillation and outperforms the classic FT method.

\subsection{Rising bubbles}
\begin{figure}[ht]
    \centering
    \includegraphics[scale=0.1]{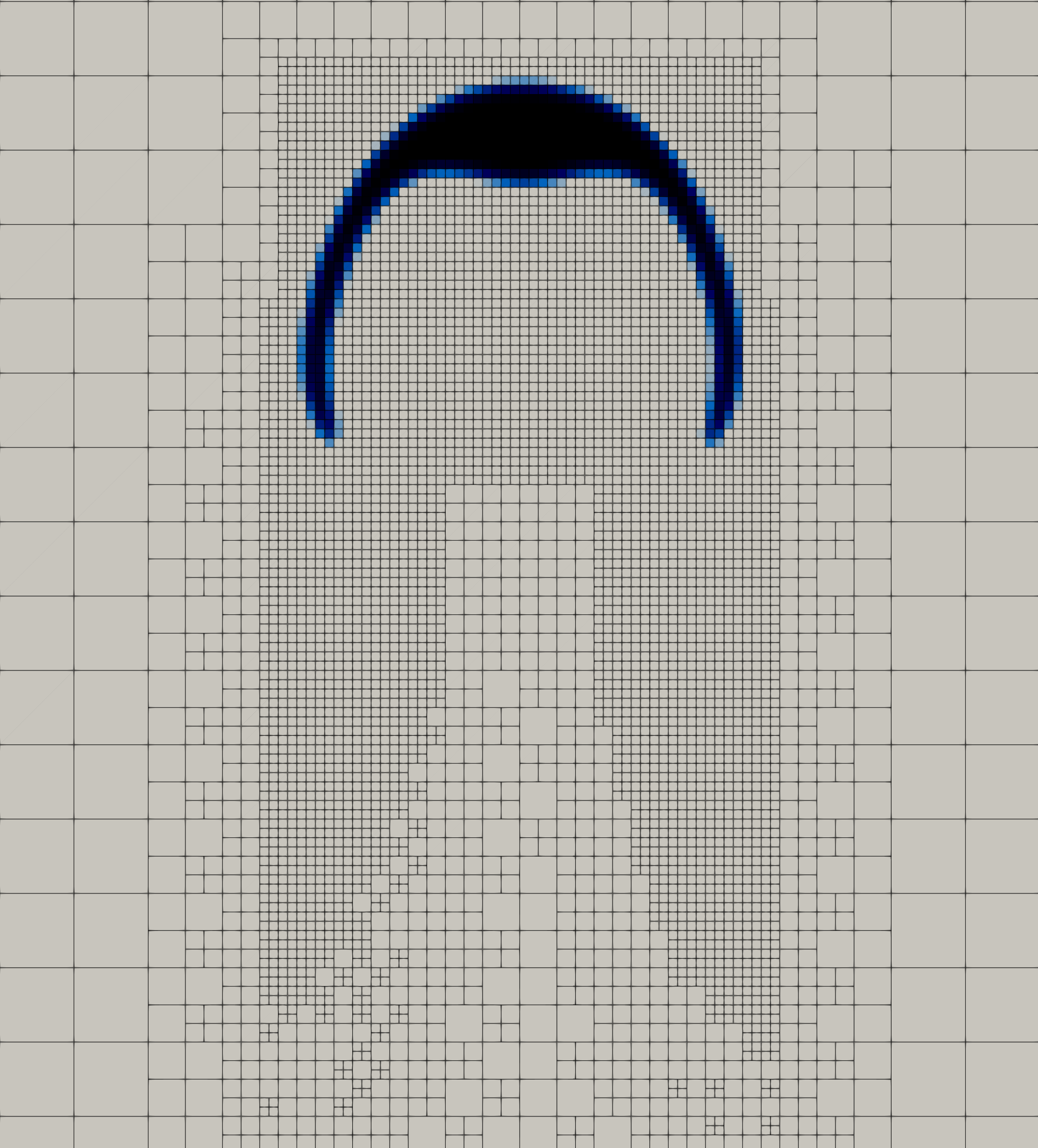}
    \caption{\reviewerRevtwo{Slice of an examplary} AMR mesh with refinement based on the front location and local flow vorticity, for the rising bubble test case.}
    \label{fig:AMR_Mesh}
\end{figure}

The results presented so far mostly rely on analytical and theoretical scenarios for validation purposes. In this section, the sharp FT method is validated and compared against experimental results of rising bubbles and the classic FT method to test the capabilities of the proposed method in a more practical and application-orientated scenario. In order to focus on the differences in the computation of the surface tension, indicator function and the velocity interpolation between classic and sharp front tracking, both methods use the same settings for remeshing as described in \cite{Gorges2022} and the local roughness smoothing for undulations removal and mesh quality. Four rising bubbles from the experiments of \citet{Bhaga1981} are considered and the numerical settings are adopted from the work of \citet{Pivello2014} and summarized in Table \ref{tab:RisingBubbleSetups}. The domain has periodic side walls, a no-slip bottom wall and a pressure-outlet on the top. An adaptive mesh refinement (AMR) with 4 refinement levels is used. The refinement is based on the front location and local flow vorticity and the highest refinement level is set to match the $d/\Delta x$ ratio in Table \ref{tab:RisingBubbleSetups}. The desired edge length for the front mesh remeshing is set to $3/4$ of the highest AMR refinement level. An example of the AMR mesh is shown in Figure \ref{fig:AMR_Mesh}.

\begin{table}
    \centering
    \caption{Parameters for the four considered rising bubbles adopted from \cite{Bhaga1981, Pivello2014}.}
    \label{tab:RisingBubbleSetups}
    \begin{tabular}{ p{3.5cm}||p{2cm}|p{2cm}|p{2cm}|p{2cm}  }
 Settings & Case 1 & Case 2 & Case 3 & Case 4\\
 \hline
 \hline
 Reynolds number & 0.078 & 7.77 & 18.3 & 55.3\\
 \hline
 Bond number & 8.67 & 243 & 339 & 32.2\\
 \hline
 Morton number & 711 & 266 & 43.1 & \num{8.2e-4}\\
 \hline
 $d_\mathrm{b} / \Delta x$ & \multicolumn{4}{c}{32}\\
 \hline
 $D / d_\mathrm{b}$ & \multicolumn{4}{c}{10}\\
 \hline
 $\rho_\mathrm{c} / \rho_\mathrm{d}$ & \multicolumn{4}{c}{1000}\\
 \hline
 $\mu_\mathrm{c} / \mu_\mathrm{d}$ & \multicolumn{4}{c}{100}
\end{tabular}
\end{table}

\begin{figure}
    \centering
    \includegraphics[scale=1.0]{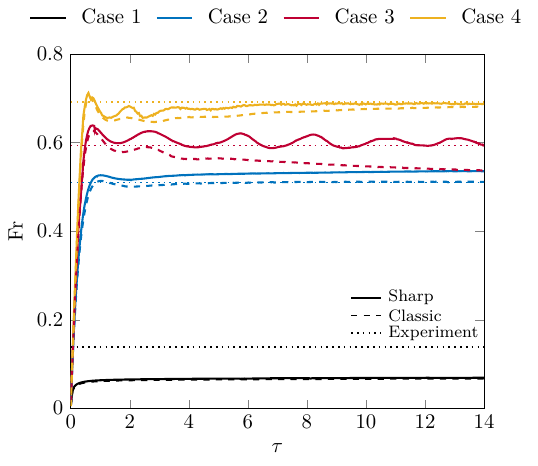}
    \caption{\reviewer{Temporal evolution of the Froude number over the dimensionless time of the rising bubbles obtained using the classic FT method and the proposed sharp FT method. The terminal Froude number in the experiments of \citet{Bhaga1981} is given as reference.}}
    \label{fig:RisingBubble_Froude}
\end{figure}

\begin{figure}
	\centering
	\subcaptionbox{Case 1: classic\label{fig:RB_C1_C}}{\includegraphics[width=3cm]{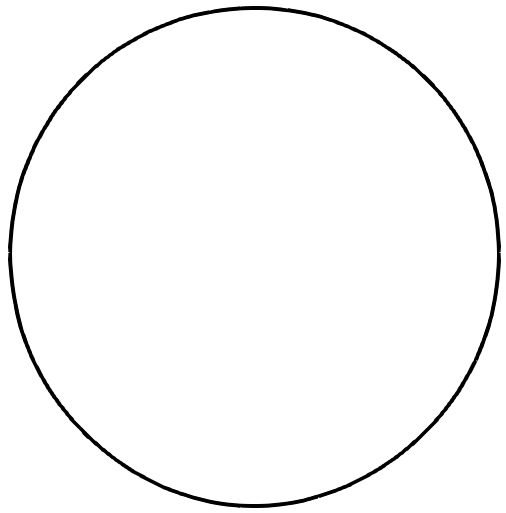}}
	\subcaptionbox{Case 1: sharp\label{fig:RB_C1_S}}{\includegraphics[width=3cm]{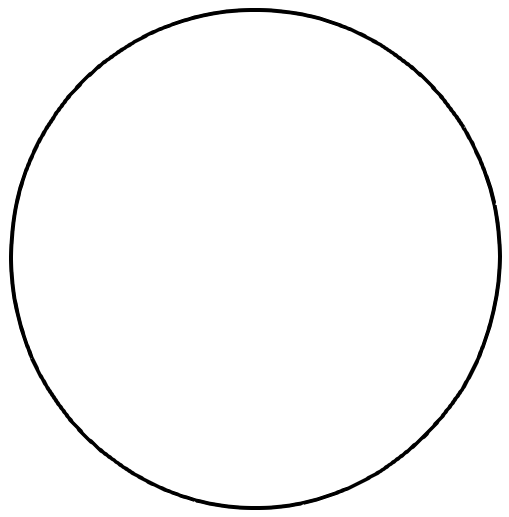}}\\
    \subcaptionbox{Case 2: classic\label{fig:RB_C2_C}}{\includegraphics[width=3cm]{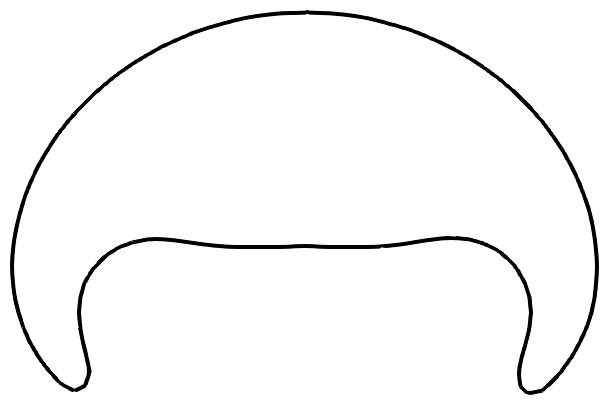}}
	\subcaptionbox{Case 2: sharp\label{fig:RB_C2_S}}{\includegraphics[width=3cm]{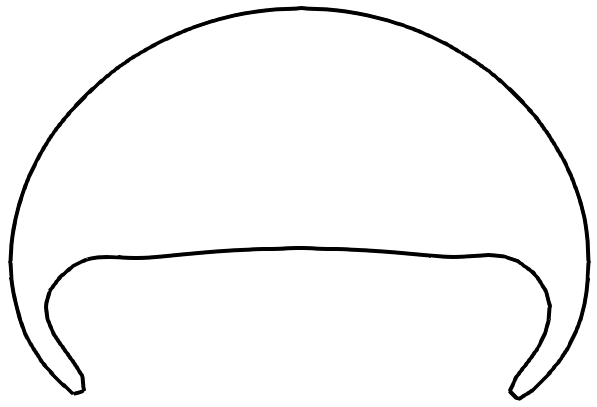}}\\
    \subcaptionbox{Case 3: classic\label{fig:RB_C3_C}}{\includegraphics[width=3cm]{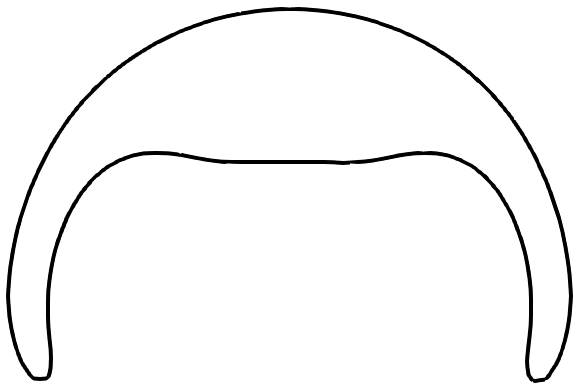}}
	\subcaptionbox{Case 3: sharp\label{fig:RB_C3_S}}{\includegraphics[width=3cm]{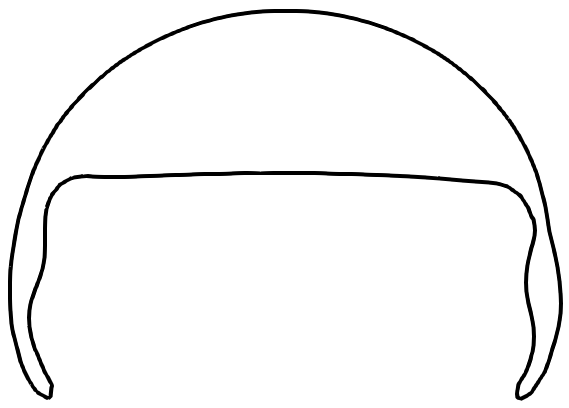}}\\
    \subcaptionbox{Case 4: classic\label{fig:RB_C4_C}}{\includegraphics[width=3cm]{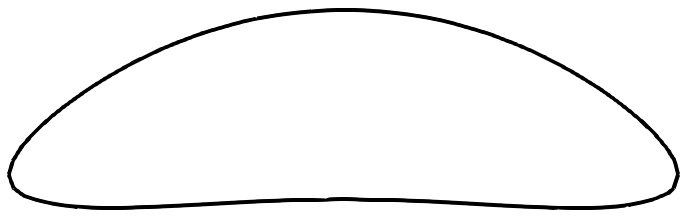}}
	\subcaptionbox{Case 4: sharp\label{fig:RB_C4_S}}{\includegraphics[width=3cm]{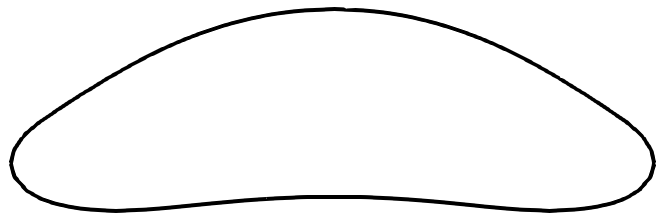}}\\
    \caption{Interface iso-contour for the classic and sharp front tracking for the different rising bubble test cases. \reviewerRevtwo{The figures show a 2D slice of the fully 3D simulations.}}
	\label{fig:RisingBubbles}
\end{figure}

\reviewer{The terminal Reynolds number $\mathrm{Re} = \rho_{\mathrm{f}} d u_{\mathrm{T}} / \mu_{\mathrm{f}}$ is based on the initial bubble diameter and the average terminal rise velocity. The Froude number is defined as $\mathrm{Fr}=u / (g d)$ and the dimensionless time is defined as $\tau = t \, \sqrt{g/d}$, with $g$ being the gravitational acceleration, $u$ the rise velocity, $t$ the time and $d$ the initial bubble diameter. Figure \ref{fig:RisingBubble_Froude} shows the temporal evolution of the Froude number for the four rising bubble test cases obtained using the classic front-tracking method and the proposed sharp-front tracking method, alongside the terminal Froude number of the experimental results of \citet{Bhaga1981}. The proposed sharp front tracking leads to overall slightly higher rise velocities compared to the classic front tracking. Especially for case 3, the rise velocity for the sharp front-tracking method oscillates around a constant rise velocity, whereas for the classic approach the rise velocity decreases. Table \ref{tab:RisingBubbleReynoldsnumbers} shows the terminal Reynolds numbers obtained with the sharp front-tracking method, the classic front-tracking method and the experimental results of \citet{Bhaga1981} for the four considered test cases. Compared to classic front tracking, the proposed sharp front tracking leads to higher terminal Reynolds numbers for all considered rising bubbles and leads to slightly smaller errors in the terminal Reynolds number compared to the experimental results, except for case 2.}
Considering that \citet{Bhaga1981} estimated their measurement error of the rise velocity to be $5\, \%$, the results obtained with the sharp front-tracking method for cases 2-4 are in agreement with the experimental measurements.
The relatively large errors for case 1 with both front-tracking methods were already explained by \cite{Hua2008}: for such low rise velocities the relative error will be high even for small absolute errors and wall effects are more significant at low Reynolds numbers. 
In \cite{Gorges2022}, the difference between the smooth Peskin velocity interpolation and the sharp divergence-preserving velocity interpolation already showed that the divergence-preserving velocity interpolation leads to slightly larger terminal Reynolds numbers, which is in agreement with the present results. The different computation methods for the surface tension force and the difference in interface thickness for the indicator function are most evident in regions of high curvature, such as the skirts of the bubbles of cases 2 and 3, shown in Figure \ref{fig:RisingBubbles}. The skirts of the bubbles predicted by the sharp front-tracking method are thinner compared to those predicted by the classic front-tracking method, which also has an influence on the wake behind the bubble and, in turn, the rise velocity. \reviewer{Overall, the iso-contours of the sharp front tracking method in Figure \ref{fig:RisingBubbles} are in very good agreement with the results of the classic front tracking method, which has been validated extensively in our previous works \cite{Gorges2022, Gorges2023} and is in excellent agreement with the literature \cite{Pivello2012, Pivello2014, Hua2008} and the experimental results of \citet{Bhaga1981}.}

\begin{table}
    \centering
    \caption{Terminal Reynolds numbers of the rising bubbles obtained using the classic FT method, the proposed sharp FT method and measured in the experiments of \citet{Bhaga1981}.}
    \label{tab:RisingBubbleReynoldsnumbers}
    \begin{tabular}{c||l|l|l}
      & \multicolumn{3}{c}{Re}\\
     \hline
     Case & Sharp FT & Classic FT & \citet{Bhaga1981}\\
     \hline
     \hline
      1 & 0.068 & 0.066 & 0.078\\
      2 & 8.15 & 7.79 & 7.77\\
      3 & 18.7 & 16.5 & 18.3\\
      4 & 55.0  & 54.5 & 55.3\\
    \end{tabular}
\end{table}

\section{Conclusions}
\label{sec:conclusions}
The sharp front-tracking method has been proposed with the aim of avoiding large interpolation kernels for the interpolation of the velocity and the interface information, and to provide a sharp interface representation on the fluid mesh in order to improve the accuracy of front tracking and extend its range of applications. To this end, the proposed sharp front tracking utilizes the compact divergence-preserving interpolation method \cite{Gorges2022}, a piecewise parabolic interface reconstruction (PPIC) based on the volumetric fit of \citet{Jibben2019}, the machine accurate and geometrically robust polyhedron intersection algorithm \cite{Evrard2023} for the reconstruction of the sharp indicator function and for the mean curvature and the normal vector computation of the PPIC, and the CSF method of \citet{Brackbill1992} for the computation of the source term representing surface tension on the fluid mesh. In addition, a novel local roughness smoothing algorithm has been proposed to suppress undulation in the front mesh as a result of parasitic currents or subgrid velocity variations. For that, a watermark detection algorithm \cite{Karni2000, Corsini2007} and the smoothing algorithm of \citet{Kuprat2001} are applied.

The proposed method has been validated, tested, and compared to the classic FT method \cite{Tryggvason2001} for various canonical interfacial flow test cases, including a stationary and translating droplet in Laplace equilibrium, oscillating droplets and rising bubbles. The presented results demonstrate that the sharp FT method is superior to the classic FT method in all tested scenarios, with the advantage of sharp interpolation kernels and a sharp representation of the interface on the fluid mesh. For the stationary and translating Laplace equilibrium test case, the proposed method leads to approximately two orders of magnitude smaller parasitic currents and a better convergence behaviour compared to the classic FT method. Furthermore, for both the Laplace equilibrium and the oscillating droplet, parameter ranges can be simulated with the proposed sharp FT method that cannot be simulated with the classic FT method in a stable manner. A comparison of the results for the Laplace equilibrium test cases also shows that the presented method is as or even more accurate than interface capturing and state-of-the-art hybrid level-set/FT methods \cite{Abadie2015, Tolle2020}. Due to the sharp interface representation on the fluid mesh, the sharp FT method is generally more sensitive to distortions for which the proposed local roughness smoothing provides a remedy and further extends the parameter range that can be simulated.

In the context of interfacial flow modelling, the proposed sharp FT method improves the overall accuracy and robustness of the front tracking method by reducing parasitic currents and providing a sharp interface representation with sharp velocity interpolation. This opens new possibilities to further extend the use of front tracking to problems where a sharp interface representation may have a significant impact on the results, such as microfluidics, fluid-fluid and fluid-structure interactions.

\section*{Data Availability Statement}
\noindent The data that support the findings of this study are reproducible
and data is openly available in the repository with DOI
10.5281/zenodo.14749559, available at \href{https://doi.org/10.5281/zenodo.14749559}{https://doi.org/10.5281/zenodo.14749559}

\section*{Acknowledgements}
\noindent This project has received funding from the Deutsche Forschungsgemeinschaft (DFG, German Research Foundation), grant number 420239128, and from the European Unions's Horizon 2020 research and innovation programme under the Marie Sk\l odowska-Curie grant agreement No 101026017. This work was supported by the US Department of Energy through the Los Alamos National Laboratory. Los Alamos National Laboratory is operated by Triad National Security, LLC, for the National Nuclear Security Administration of U.S. Department of Energy (Contract No. 89233218CNA000001).

\appendix

\Editor{
\section{Relation of curvature to the roughness metric}
\label{AppendixA}
By computing the deviation between adjacent triangle normals, the roughness metric is essentially a measure of local curvature. The subsequent averaging and thresholding steps translate these local roughness measures into actionable metrics for further processing, such as smoothing operations.
The edge roughness is given by
\begin{equation}
    \mathcal{R}_e = 1 - (\mathbf{n}_1 \cdot \mathbf{n}_2)
    = 1 - \cos{(\Theta)},
\end{equation}
where $\Theta$ is the dihedral angle between the triangle normal vectors.
The dihedral angle is, thus, defined as 
\begin{equation}
    \label{eq:1}
    \Theta = \arccos{(1 - \mathcal{R}_e)},
\end{equation}
which, for a small perturbation along a curved surface, can be approximated by
\begin{equation}
    \label{eq:2}
    \Theta \approx \kappa \cdot d ,
\end{equation}
with $d$ being the distance between the triangle centers, or triangle normal vectors, respectively. This distance scales with the triangle edge length $l_e$ and the fluid mesh resolution $\Delta x$ based on the predefined settings of the simulation setup.

Substituting Eq.~\eqref{eq:2} into Eq.~\eqref{eq:1} and reformulating, we obtain
\begin{equation}
    \label{eq:3}
    \kappa \approx \frac{\arccos{(1 - \mathcal{R}_e)}}{d}.
\end{equation}
Assuming that, locally, the vertex roughness is dominated by the maximum edge roughness and therefore neglecting the neighbourhood averaging/smoothing, we obtain
\begin{equation}
    \label{eq:4}
    \kappa \approx \frac{\arccos{(1 - \mathcal{R}_{x_i})}}{d} .
\end{equation}
To find the largest physical curvature $\kappa_{\mathrm{max}}$ that does not trigger roughness smoothing, we assume
\begin{equation}
    \kappa_{\mathrm{max}} \approx \frac{\arccos{(1 - \mathcal{R}_{x_i}^{\mathrm{thr}})}}{d} .
\end{equation}
If the threshold is chosen such that $\mathcal{R}_{x_i}^{\mathrm{thr}} = \alpha \mathcal{R}_{x_i}^{0}$, with $\alpha$ being a factor (e.g. 1.2), $\kappa_{\mathrm{max}}$ follows as
\begin{equation}
    \kappa_{\mathrm{max}} \approx \frac{\arccos{(1 - \alpha \mathcal{R}_{x_i}^{0})}}{d} .
\end{equation} 

We have tested the accuracy of Eqs.~\eqref{eq:3} and \eqref{eq:4} against the analytical curvature of a sphere. The average errors are below 1\% and 2\%, respectively. Furthermore, the error remains approximately constant as the mesh is refined or coarsened, since the roughness metric is based on local geometric relationships that are inherently scale-invariant. As a result, the method exhibits a consistent (and already small) error that is intrinsic to its geometric definition rather than being dominated by discretization errors.

Finally, we note that the derivation above follows the mean curvature definition 
\begin{equation}
    H = \frac{1}{2} (\kappa_1 + \kappa_2),
\end{equation}
where $\kappa_1$ and $\kappa_2$ are the principal curvatures. However, in computational fluid dynamics (CFD), curvature is often defined as $\kappa = 2H = {2}/{R}$ for a sphere of radius $R$. Consequently, it may be necessary to include a factor of $2$ in the equations depending on the convention used in the application:
\begin{equation}
    \kappa_{\mathrm{max}} \approx 2 \, \frac{\arccos{(1 - \alpha \mathcal{R}_{x_i}^{0})}}{d} .
\end{equation} 
}


\end{document}